\documentclass{emulateapj}
\usepackage{times}

\newcommand{\ms}{preprint} %{preprint} %{ms} %**********************************************
\usepackage{ifthen}

\newcommand{\gray}{$\gamma$-ray}
\newcommand{\grays}{$\gamma$-rays}

\newcommand{\hii}{H~{\sc ii}}

\newcommand{\PAO}{Auger}% Observatory}
\newcommand{\AC}{Auger Collaboration}
\newcommand{\thetaav}{\langle\theta_d\rangle}

\newcommand{\pubbook}[6]{#1, #6, #2, #3, #4, #5}
\newcommand{\pubjournal}[6]{#1 #5, #2, #3, #4}
\newcommand{\pubjournala}[6]{#1 #5, #2, #4}

\newcommand{\pubjournalb}[6]{#1 #5, #2  #4}

\newcommand{\nphysb}{Nucl.\ Phys.\ B}
\newcommand{\nimA}{Nucl.\ Instr.\ Meth.\ Phys.\ Res.\ A}

\newcommand{\icrc}{Int.\ Cosmic Ray Conf.\ }

\newcommand{\app}{APh}

\ifthenelse{\equal{\ms}{preprint}}%**********************************************
{\newcommand{\jcap}{JCAP}}{}
\shorttitle{Association of UHECRs with AGN}
\shortauthors{Moskalenko et al.}

\begin{document}

\title{
On the Possible Association of Ultra High Energy Cosmic Rays with Nearby Active Galaxies
}

\author{Igor V. Moskalenko\altaffilmark{1} and \L{}ukasz Stawarz\altaffilmark{2}}
\affil{
Kavli Institute for Particle Astrophysics and Cosmology,
Stanford University, Stanford, CA 94309
\email{imos@stanford.edu, stawarz@slac.stanford.edu}}
\altaffiltext{1}{Also 
Hansen Experimental Physics Laboratory, 
Stanford University, Stanford, CA 94305
}
\altaffiltext{2}{Also
Astronomical Observatory, Jagiellonian University, ul. Orla 171, 30-244 Krak\'ow, Poland
}
%\altaffiltext{1}{Also 
%Kavli Institute for Particle Astrophysics and Cosmology,
%Stanford University, Stanford, CA 94309}

\author{Troy A. Porter}
\affil{
  Santa Cruz Institute for Particle Physics,
  University of California, Santa Cruz, CA 95064
\email{tporter@scipp.ucsc.edu}}

\and

\author{Chi C. Cheung}
\affil{NASA Goddard Space Flight Center,
Astrophysics Science Division, Code 661,
Greenbelt, MD 20771
\email{Teddy.Cheung@nasa.gov}}

\begin{abstract}
Data collected by the Pierre Auger Observatory (\PAO) provide evidence for 
anisotropy in the arrival directions of cosmic rays (CRs) 
with energies $>$57 EeV that suggests a correlation with the 
positions of active galactic nuclei (AGN) located
within $\sim$75 Mpc 
and $3.2^{\circ}$ of the arrival directions.
This analysis, however, does not take into account AGN morphology.
A detailed study of the sample of AGN whose positions are located
within $3.2^{\circ}$ of the CR events
(and extending our analysis out to $\sim$150 Mpc)
shows that most of them are classified as Seyfert 2
and low-ionization nuclear
emission-line region (LINER) galaxies whose properties do not
differ substantially from other local AGN of the same types. 
Therefore, if the production of the highest 
energy CRs is 
persistent 
in nature, i.e., operates in a single object 
on long ($\ga$ Myr) timescales, the claimed correlation
between the CR events observed by \PAO{}
and local active galaxies should be
considered as resulting from a chance coincidence.
Additionally, most of the selected sources do not show significant jet
activity, and hence, 
in most conservative scenarios,
there are no reasons for expecting
them to accelerate CRs up to the highest energies, $\sim$$10^{20}$ eV.
If the extragalactic magnetic fields and the sources 
of these CRs
are coupled with matter, it is possible that the deflection angle
is larger than expected in the case of a uniform source distribution
due to effectively larger fields.
A future analysis has to take into account AGN morphology and may yield
a correlation with a larger deflection angle and/or more distant sources.
We further argue that the nearby radio galaxy NGC 5128 (Cen~A) alone
could be associated with at least 4 events due to its large radio extent,
and PKS~1343--60 (Cen~B), another nearby radio galaxy, can be associated 
with more than 1 event due to its proximity to 
the Galactic plane and, correspondingly, the stronger Galactic magnetic 
field the ultra high energy CRs (UHECRs) encounter during propagation to the Earth.
If the UHECRs associated with these events are 
indeed accelerated by Cen~A and Cen~B, their deflection angles may
provide information on the structure of the magnetic field in the direction 
of these putative sources.
Future \gray{} observations (by, e.g., 
\emph{Fermi Gamma-ray Space Telescope} [Fermi] formerly
\emph{Gamma-Ray Large Area Space Telescope}
[GLAST], 
and
\emph{High Energy Stereoscopic System} [HESS] 
in the Southern hemisphere) 
may provide additional 
clues to the nature of the accelerators of the UHECRs in the local Universe.

\end{abstract}


\section{Introduction}\label{intro}
%##############################################################################
The spectrum, origin, and composition of CRs at the highest energies 
($\ge 10^{18}$ eV $\equiv 1$ EeV; hereafter UHECRs)
has been a puzzle since their discovery almost 50 years ago
\citep[for a review see][]{NW2000,Cronin2005}.
The isotropy of the arrival directions of UHECRs above $10^{18}$ eV suggests
an extragalactic origin, though Galactic objects such as 
fast rotating neutron stars with ultrastrong magnetic fields 
(``magnetars'')
are capable
of accelerating particles up to $\sim$$10^{20}$ eV \citep{Hillas1984}.
The UHECR energy losses due to photopion production
on the cosmic microwave background (CMB) -- the so-called GZK effect 
\citep{Greisen1966,ZK1966} -- mean that the sources of the highest energy 
particles should be cosmologically close, within $\sim$$100$ Mpc.
However, observations of the spectrum at these high energies are extremely 
challenging due to the low overall event 
rate ($\sim$1 per km$^2$ per century at $\sim$$10^{20}$ eV).
Fine structure in the CR spectrum above $10^{18}$ eV has been predicted 
\citep[e.g.,][]{Berezinsky2006} --
the GZK cutoff, a pile-up bump, and a dip due to photoproduction of pairs --
and indeed found in experiments, but the interpretation is not straightforward 
due to the large uncertainty in the source distribution,
their injection spectrum, and CR chemical composition.

Quite a few extragalactic objects
discussed in the literature are potentially
capable of accelerating UHECRs.
Among those are shocks from the epoch of the large scale structure formation,
\gray{} bursts, galaxy clusters, 
AGN (in particular, those AGN with powerful jets) and powerful AGN flares
\citep{Farrar2008}, the lobes of giant 
radio galaxies \citep{Stanev2007}, 
and newly born magnetars in other galaxies \citep{ghi08};
various exotic top-down scenarios
have also been discussed but seem 
unlikely \citep[e.g.,][]{Abraham2007a,Abraham2008b}.
Extragalactic jets and their extended radio lobes have been proposed as 
one of the most likely acceleration sites of UHECRs
\citep[e.g.,][and references therein]{BS1987,ost02,Dermer2007}.
In this paper we, therefore, will follow the ``jet paradigm'',
i.e.\ assume that UHECR particle acceleration
takes place in jets or is associated with them. 
However, not all AGN have jets, and for those that do the jet properties may 
differ 
substantially between different classes/types
(see a discussion in \S\ref{agn_ph}).

There are also quite a few ``known unknowns'' which affect the distribution
of the arrival directions of UHECRs and make the association with 
particular classes of objects a non-trivial task.
Among these
are the source distribution, the structure of extragalactic and 
Galactic magnetic fields, the nature of sources 
(transient vs.\ steady), and the energy spectrum and chemical composition
at injection 
(we discuss these topics in \S\ref{propagation}).

The \AC{}
%\PAO{}
has reported significant evidence for anisotropy 
in the arrival directions of UHECRs \citep{Abraham2007b,Abraham2008a}.
The anisotropy signal suggests a correlation of the events 
with AGN listed in the 
\citet{VC2006} catalog with distances less
than $\sim$100 Mpc, though other sources with a similar distribution
are not ruled out\footnote
{Note that the \AC{}
%\PAO{}
acknowledges the incompleteness of the
catalog and does not make any serious claims concerning source classes.
}. 
The maximum correlation has been found for AGN with redshift
$z\le0.017$ (corresponding distance $D\le71$ Mpc), angular separation
$\theta\le3.2^\circ$, and events with energy above $\sim$57 EeV. 
The list of events with 
energy in excess of 57 EeV consists of 27 events, and 20 of these correlate
with the AGN from the \citet{VC2006} catalog.

In this paper we discuss the possible association between UHECRs and AGN 
based on a detailed analysis of a selected sample of nearby active
galaxies contained within the $\theta\le3.2^{\circ}$ search radius of the UHECR 
events
detected by \PAO, 
drawn from the \citet{VC2006} catalog, with additional AGN taken from 
the NASA/IPAC Extragalactic Database (NED). 
Our study of the properties of the selected sample of AGN, described here in detail,
shows that it consists 
predominantly of low-luminosity sources such as Seyfert galaxies and LINERs, 
and a handful of radio galaxies.

In \S\ref{agn_ph} we give a brief introduction (not intended as a comprehensive review)
into the AGN phenomenology targeting non-expert readers; those familiar 
with the AGN background material could safely skip it.
In \S\ref{agn_sample} we discuss a sample of local AGN selected using exactly the same 
criteria ($\theta\le3.2^{\circ}$ and $z \leq 0.018$) for each UHECR event 
that correspond to the maximum significance of the correlation 
reported by the \AC.
%\PAO.
We extend the search out to redshifts $z \leq 0.037$ which doubles the distance
to 150 Mpc, and effectively covers the range of horizons for super-GZK particles
\citep{Harari2006,Abraham2008a}.
With such a sample, 
we investigate the properties of the listed 54 active galaxies. In particular, 
we address the two following issues: (i) Are the selected objects different in any 
respect from the other numerous local AGNs? (ii) Are they in general able to 
accelerate cosmic rays to ultra high energies?
Based on the selection criteria, we identify
four powerful radio galaxies which could be potential sources of,
at least, 7 out of 27 reported UHECR events.
We complete with the discussion of the propagation of 
UHECRs from the sources to the observer (\S\ref{propagation}), and argue that the
correlation radius between the arrival direction of UHECRs and
the sources could be significantly larger than $\theta\le3.2^{\circ}$
found by the \AC{}
%\PAO{}
and
depends on the UHECR source distribution and the assumed model of 
intergalactic and Galactic magnetic fields. In
\S\ref{conclusion} we summarize our reasoning that the observed correlation 
is most probably chance coincidence given the high density of low
power AGN in the local universe.

\section{Phenomenology of local AGN}\label{agn_ph}
%##############################################################################
A number of weak AGN in the local Universe can be observed and even resolved
due to their proximity which allows a broad range of AGN activity to be 
studied, while only bright ones can be seen at large distances.
At the current state of our knowledge, however,
not all AGN types could be recognized as potential sources of UHECR.
A brief introduction into the types of AGN activity is, therefore,
warranted as a basis for further discussion.

\subsection{Nuclei}\label{nuclei}
%##############################################################################
Supermassive black holes 
($\mathcal{M}_{\rm BH} \sim 10^6 \mathcal{M}_{\odot} -
10^9 \mathcal{M}_{\odot}$) are found in a number of galaxies,
and there is growing consensus that all galaxies contain accreting
black holes at their centers. 
If the emission of the
accreting/circumnuclear matter is pronounced, the galaxy is classified
as active. 
Such emission includes a non-stellar blue continuum due to
the accretion disk and strong, but narrow, forbidden emission lines
resulting from photoionization of the surrounding gas by the disk
radiation. 
It has been shown 
that about $85\%$ of all galaxies
(and, in particular, almost all nearby late-type galaxies) possess
detectable emission-line nuclei, which therefore could be classified
as AGN \citep{ho08}. 
In most cases, however, the
nuclear luminosity is very weak relative to the stellar radiative
output of the host galaxy\footnote{If the nuclear emission outshines
the starlight of the host galaxy, the source is classified as a
quasar.}. 
These nearby low-luminosity AGN can be divided into Seyfert
galaxies ($\gtrsim$$10\%$ of local AGN) and 
LINERs ($\sim$$60\%$), depending on the
ionization level and intensity of the emission lines\footnote{We note
that the starburst nuclei of many galaxies, as well as their \hii{}
regions, also possess ionization lines, which are, however, weaker than
the ones observed in Seyferts or LINERs, and are photoionized
exclusively by the starburst activity.}. 
For more details we refer
the reader to the recent review by \citet{ho08}.

The rest
of the local AGN assemblage is populated by radio galaxies \citep[mostly 
low-power ones of the FR~I type;][]{fr74} and BL Lacertae objects (BL Lacs). 
In these sources, the radiative output of the accreting/circumnuclear gas 
(which may,  
spectroscopically, resemble closely the Seyfert or LINER types) 
is dominated, 
or at least substantially modified, by a broad-band non-thermal 
emission produced
by relativistic jets emanating from the active centers. 
Also, 
radio galaxies and BL Lacs are hosted almost
exclusively by giant elliptical and S0 galaxies, while the majority of 
Seyferts and LINERs ($\gtrsim$$90\%$) are associated with spiral
galaxies 
\citep[e.g.,][]{1997ApJ...479..642B,1998ApJS..117...25M,1999ApJS..122...81M}. 
Therefore, 
Seyferts/LINERs and FR~Is/BL Lacs
are \emph{morphologically different} and this
may have important 
consequences/implications for their capability of accelerating particles 
to the highest observed energies, 
as discussed in the next section.

Some fraction of AGN exhibit in addition broad permitted emission lines 
(${\rm FWHM} \sim 10^3-10^4$\,km\,s$^{-1}$) in their spectra;
the presence/absence of such lines divides further the AGN population into
type~1/type~2 classes. 
In the framework of the ``AGN unification scheme'',
these two classes are intrinsically identical, and differ only in
the orientation of the accretion disk symmetry axis to the line of sight
\citep{ant93}. 
Namely, type~1 AGN are 
thought to be predominantly 
observed at
small inclination angles (roughly $<$$40^{\circ}$), while type~2 AGN are
viewed at larger angles through a high-column-density
gas and dust concentrated at pc-scale distances from the active center
in a torus-like structure (co-planar with the inner accretion disks). 
As
a result, the disk continuum and broad emission lines produced very
close to supermassive black holes are heavily obscured in type~2 AGN,
but not in type~1 sources. 
The narrow line emission, originating at
larger distances ($>$~pc) from the active center, is not subjected to
strong obscuration. 

Such a unifying picture is supported by several
observational findings, including detections of strong mid-infrared (MIR)
emission due to the dusty obscuring tori in basically all different classes and
types of AGN. 
In addition to the MIR/optical/UV non-stellar nuclear
radiation, Seyferts and LINERs show characteristic X-ray emission
extending from $0.1$ keV up to $100$ keV photon energies with
photon indices $\Gamma_X \lesssim 2.0$, 
which is probably
produced within the hot coronae of the accretion disks \citep[see,
e.g.,][]{sve96,Zdziarski1999}. 
The low-energy ($<2$ keV) segment of this continuum
is typically obscured in type~2 sources, which agrees with the
unification scheme. 
We also note that in the framework of the
unification paradigm, all BL Lacs are 
considered
to be simply
low-luminosity radio galaxies (FR~Is) viewed at very small viewing
angles ($\lesssim 10^{\circ}$) to the jet axis.

\subsection{Jets}\label{jets}
%##############################################################################

Acceleration of particles to ultrarelativistic energies in
AGN is observationally confirmed (through detection of the intense non-thermal emission) to
be associated with collimated fast jets, which are produced by the active black
hole/accretion disk systems\footnote{Note, that there is no observational evidence for a
significant population of non-thermal particles in AGN accretion disks and disk coronae.
In fact, hard X-ray/soft $\gamma$-ray emission observed from Seyferts is best modeled
assuming a thermal population of electrons, albeit characterized by high temperatures of
the order of $100$ keV.}, or in shocks created by a propagating jet.
The strong magnetization, extremely low density,
non-stationary relativistic and supersonic flow pattern, and finally
turbulent character of such jets \citep[see, e.g., the review
by][]{beg84}, is expected to result in the formation of non-thermal
particle energy distributions extending up to the highest accessible
energies. 
Indeed, the very high energy $\gamma$-ray emission detected
from blazars\footnote{Blazar class includes BL Lacertae objects and
radio loud quasars with flat spectrum radio cores.}, demonstrates the
ability of the nuclear (sub-pc scale) relativistic AGN outflows to efficiently
accelerate electrons up to $1-100$ TeV energies
\citep[e.g.,][]{cel08}. 
In the case of large (kpc)-scale jets, as
observed in powerful radio galaxies and radio-loud quasars, it has been 
speculated that proton energies in the $10-100$ EeV
range can be reached as well \citep[e.g.,][]{ost98,lyu07}. 

Not all AGN are jetted.
Also, it seems that only some fraction of nuclear jets
can remain relativistic and well-collimated during the propagation
through the dense environments of the central parts of host
galaxies. 
These issues are particularly relevant for Seyferts and
LINERs, which differ substantially from the bona fide jet sources like
radio galaxies and radio-loud quasars, lacking in particular
well-defined large-scale jet structures.
We also note in this
context, that no galaxy classified as Seyfert or LINER has been
detected so far at $\gamma$-ray ($>$MeV) energies\footnote{This may 
change with the next generation of $\gamma$-ray satellites
(\emph{Fermi}/Large Area Telescope [\emph{Fermi}/LAT]) 
and ground-based Cherenkov Telescopes (such as the
next phases of HESS, \emph{Major Atmospheric Gamma-ray Imaging Cherenkov telescope} [MAGIC], 
or \emph{Very Energetic Radiation Imaging Telescope Array System} [VERITAS]).}.

The jet production and acceleration efficiency may
depend on the
\emph{morphology} of the host galaxy. 
\citet{sik07} proposed that the difference in the 
efficiency of the jet production between spiral-hosted AGN (Seyfert 
galaxies and LINERs) and elliptical-hosted ones 
(low-power FR~I radio-galaxies, their more powerful analogs classified as FR~II radio sources,
BL Lacs, quasars) may be explained if the spin of the central supermassive black 
holes, $J$, in spiral galaxies is on average lower than in elliptical ones. 
According to the so-called ``spin paradigm'', the efficiency for the production of 
relativistic jets in AGN depends on the spin of the supermassive black hole 
\citep{bland90}. This has been investigated in detail
by \citet{vol07}, who,
by means of numerical simulation,
confirmed that indeed the cosmological evolution of supermassive black holes
is expected to lead to such a host morphology-related spin dichotomy. If true,
this may have important consequence for the acceleration of CRs to the highest
observed energies. Namely, the electromotive force of the black hole embedded
in an external magnetic field $B$ (supported by the accreting matter) is 
$\Delta V \sim B \, r_g \, (J/J_{\max})$ \citep[e.g.,][]{phi83},
where $J_{\max}= c r_g \mathcal{M}_{\rm BH}$ is the maximum value of the black hole spin, 
and $r_g$ is the gravitational radius of the black hole.
Assuming further 
that the magnetic energy density close to the 
black hole event horizon 
is equal to the 
energy density of the matter accreting at the Eddington rate, 
the maximum energy a test particle 
with an electric charge $Ze$ can reach in such a potential drop is
$E_{\max} \sim Ze \, \Delta V \sim 3 \times 10^{20} Z (\mathcal{M}_{\rm BH} 
/ 10^8 \mathcal{M}_{\odot})^{1/2} (J/J_{\max})$ eV. 
If this picture is correct then 
many (most) of the local low-power AGN with $\mathcal{M}_{\rm BH} < 
10^8 \mathcal{M}_{\odot}$ and accreting at sub-Eddington rates, and are 
additionally characterized by $J \ll J_{\max}$ as expected in the case of 
the spiral-hosted Seyferts/LINERs, may not have enough potential to 
accelerate CRs to ultra high energies \citep{hop06,sik07,vol07}.

Jet activity always manifests itself as 
non-thermal
synchrotron radio emission produced by ultrarelativistic jet
electrons. 
Hence, investigation of jet properties are usually
addressed by detailed radio studies. 
In the case of local AGN, like
Seyferts and LINERs, such studies are hampered due to the low luminosities
and small sizes of the radio structures. 
Additionally, most of these
sources, and in principle all local spiral-hosted AGN, show intense
star formation activity (especially pronounced in the far-infrared [FIR]),
see, e.g., \citet{ho08}.
Such activity
is known for producing the radio-emitting outflows due to
starburst-driven superwinds which, although completely different in
origin, may in some (morphological and spectral) respects closely resemble
jet-related activity. 
Thus, care must be taken when making 
any statement regarding jet properties
in these objects with limited radio data.

The synchrotron/jet origin of the compact cores observed in local
low-luminosity AGN is supported by the direct detections of nuclear
jets in these systems \citep{ulv98,kuk99,mid04}. 
Interestingly,
observations of the proper motions of such structures indicate
sub-relativistic bulk velocities on sub-pc/pc-scales, $v < 0.25 c$
\citep{ulv99a,ulv99b,mid04}. 
Note that the observed brightness
temperatures of the radio cores in Seyferts and LINERs, as well as
their moderate year-timescale variability, do not require
relativistic beaming \citep{mun00}. 
Also, the observed one-sidedness of
nuclear jets in these systems is best explained as resulting from
free-free absorption of radio photons on the surrounding gaseous
disks, and not due to relativistic (Doppler) effects
\citep{mid04}. 
This constitutes a clear difference with the established
properties of radio galaxies and quasars. 

We also emphasize that the
distribution of nuclear jets in Seyferts and LINERs is random with
respect to the host galaxy stellar disks
\citep{nw99,pri99,kin00,sch01,the01,sch02}. 
Moreover, misalignments
between sub-pc/pc-scale jets and kpc-scale radio structures are common
in Seyferts, being much larger than those observed in other
radio-loud AGN \citep{col96,mid04,gal06,kha06}. 
The
misalignment angle distribution is flat over the whole range 
$0^{\circ}$ to $90^{\circ}$ \citep{mid04}. 
This may suggest that
kpc-scale radio structures are not powered by the jets, but rather
originate in starforming regions. 
Such an interpretation is sometimes
supported by the fact that the extended radio-emitting halos in some
Seyferts and LINERS are aligned with the galaxy disks, and that their
radio powers correlate with the FIR fluxes as observed in regular
spiral and starforming galaxies \citep{the01}. 

However, in most of the
cases ($\geq 45\%$) the observed kpc-scale radio structures of
Seyferts and LINERs do not morphologically match those of galaxy disks or
starforming regions, showing also excess radio emission over the
radio-FIR correlation \citep{gal06,kha06}. 
Thus, such structures are
believed to be powered directly by jets, which are, however,
substantially different from the ones observed in other radio-loud
AGN. 
In particular, there is growing consensus that Seyferts and
LINERs are characterized by short-lived low-power jet activity
epochs, with jet lifetimes $t_{\rm j} \lesssim
10^5$ yrs and jet kinetic luminosities $L_{\rm j} \sim
10^{41}-10^{43}$ erg s$^{-1}$, possibly triggered by minor accretion
episodes, which repeat every $\sim$$10^6$ yrs in different (random)
directions over the whole Seyfert-type activity epoch $\lesssim$ Gyr
\citep{san84,cap99,kha06,gal06}.

\subsection{Radio activity}\label{radio}
%##############################################################################

Radio surveys of 
local Seyferts and LINERs
provide further insight
into the nature of their activity and
reveal typically complex,
multi-component radio structures, consisting of compact unresolved or
slightly resolved cores (detected at centimeter wavelengths in most of these
objects), linear jet-like features (observed in $\sim$$30\%$ of the
local low-luminosity AGN), and spherical or elongated diffuse
halos/radio bubbles \citep[present in, again, $\sim$$30\%-50\%$ of the
Seyfert and LINER population;
e.g.,][]{bau93,ulv89,col96,rus96,mor99,ho01,gal06}. 
The total radio
power of these objects at centimeter wavelengths is in the range 
$\sim 10^{35}-10^{42}$ erg s$^{-1}$ \citep[with median value $L_{\rm
5\,GHz} \sim 2 \times 10^{37}$ erg s$^{-1}$;][]{ho01}, which is
typically slightly higher than the total radio luminosities of
``regular'' spiral galaxies.

For the compact cores of Seyfert and LINERs, 
previously no dependence in radio power or spectral properties on the
AGN type (type~1 vs.\ type~2) or host galaxy morphology (ellipticals
vs.\ spirals) was found \citep{ulv89,rus96,mor99,nag99}. 
However, 
more recently 
it has been suggested that the small fraction of those Seyferts
which are elliptical-hosted, and possibly also type~1 Seyferts, are
characterized by relatively stronger nuclear radio activity
\citep{the01,ulv01,nag05}. 
Nevertheless, it is established that
flat-spectrum radio cores can be present in all types of local
low-luminosity AGN, and therefore (unlike in the case of radio
galaxies and quasars) cannot be used as a good proxy for the source
inclination. 

It has been speculated that the flat-spectrum radio
cores of Seyferts may not be due to jet activity, but rather some other
processes like a nuclear starburst, or the emission of the accretion
disks/obscuring tori themselves \citep[e.g.,][]{gal97}. 
However, 
the relatively high brightness temperatures of the Seyfert radio 
cores, $10^7\ {\rm K} < T_{\rm b} < 10^9$ K, points to a synchrotron 
(and therefore jet-related) origin of the radio emission
\citep{mun00,mid04,nag05}. 
This is supported by the excess radio
emission of the local low-luminosity AGN with detected radio
cores over the radio--FIR correlation established for regular spirals
and starforming galaxies \citep{roy98,ho01} which is expected to hold if
the radio emission is exclusively/predominantly due to starforming
activity \citep{hel85,con02}. 
Also, recently found scaling relations
between the nuclear radio emission of Seyferts and LINERs and their
accretion power (approximated by either the $2-10$ keV luminosity of the
disk coronae, non-stellar optical magnitudes of the active nuclei, or
H$_\beta$ or O\ {\sc iii} line luminosities), indicate a strong
link between the radio production efficiency and the accretion disk
parameters, as expected in the case of a jet origin for the radio
emission \citep{hp01,ho02,pan06,pan07}.

\section{AGN sample}\label{agn_sample}
%##############################################################################

\placetable{Table~\ref{t1}}

\begin{deluxetable*}{cccccc}
\tabletypesize{\footnotesize}
\tablecaption{\label{t1} AGN possibly associated with UHECRs events}
\tablecolumns{6}
\tablewidth{0pt}
\tablehead{
\multicolumn{2}{c}{UHECR Event} & \multicolumn{4}{c}{AGN within $3.2^{\circ}$ search radius}\\
%\multicolumn{2}{c}{} & \multicolumn{4}{c}{}\\
Event number & Galactic Coordinates & \multicolumn{2}{r}{$0 < z \leq 0.018\ \ \ $} & \multicolumn{2}{r}{$0.018 < z \leq 0.037\ \ \ $}\\
%\multicolumn{2}{c}{} & \multicolumn{2}{c}{} & \multicolumn{2}{c}{}\\
\multicolumn{1}{c}{} & \multicolumn{1}{c}{$(\ell\,[^{\circ}], \, b\,[^{\circ}])$} & \multicolumn{1}{c}{Name} & \multicolumn{1}{c}{$\theta$ [$\arcmin$]} & \multicolumn{1}{c}{Name} & \multicolumn{1}{c}{$\theta$ [$\arcmin$]}}
\startdata
25	&	($-$21.8, 54.1)	&	NGC 5506		&	38	&	UM 653		&	136	\\
		&			&				&		&	UM 654		&	138	\\
		&			&				&		&	UM 625		&	171	\\
18	&	($-$57.2, 41.8)	&				&		&	ESO 575-IG016	&	55 \\
		&			&				&		&	{\bf  MCG-03-32-017}	&	191 \\
26	&	($-$65.1, 34.5)	&				&		&	TOLOLO 00020	&	133 \\
2	&	($-$50.8, 27.6)	&	{\bf  NGC 5140}		&	116	&	{\bf  2MASX J13230241-3452464}	&	55 \\
		&			&				&		&	{\bf  TOLOLO 00081}	&	95 \\
		&			&				&		&	{\bf  ESO 444-G018}	&	144 \\
21	&	($-$109.4, 23.8)	&	{\bf  NGC 2907}		&	94	&				&		\\
		&			&	NGC 2989		&	182	&				&		\\
20	&	($-$51.4, 19.2)	&	NGC 5128 (Cen~A)$^{\star}$	&	54	&				&		\\
17	&	($-$51.2, 17.2)	&	NGC 5128 (Cen~A)$^{\star}$	&	139	&				&		\\
		&			&	{\bf  NGC 5064}$^{\star}$	&	162	&				&		 \\
		&			&	NGC 5244		&	164		&				&		\\
8	&	($-$52.8, 14.1)	&	{\bf  NGC 5064}$^{\star}$	&	47	&				&		 \\
		&			&	{\bf  IRAS 13028-4909}	&	118	&				&		\\
		&			&	NGC 4945		&	121	&				&		\\
5	&	($-$34.4, 13.0)	&	IC 4518A		&	66	&				&		\\
1	&	(15.4, 8.4)		&				&		&	{\bf  1RXS J174155.3-121157}	&	106 \\
		&			&				&		&	HB91 1739-126	&	122 \\
14	&	($-$52.3, 7.3)	&				&		&	WKK 2031		&	83	\\
23	&	($-$41.7, 5.9)	&		{\bf  WKK 4374}	&	167	&				&		 \\
3	&	($-$49.6, 1.7)	&	DZOA 4653-11	&	40	&				&		\\
		&			&	{\bf  PKS~1343-60 (Cen~B)}		&	41	&				&		\\
27	&	($-$125.2, $-$7.7)	&				&		&				&		\\
11	&	($-$103.7, $-$10.3)	&				&		&				&		\\
13	&	($-$27.6, $-$16.5)	&	ESO 139-G012$^{\star}$	&	109	&				&		\\
		&			&	{\bf  AM 1754-634 NED03}$^{\star}$	&	191	&				&		\\
4	&	($-$27.7, $-$17.0)	&	ESO 139-G012$^{\star}$	&	139	&				&		\\
		&			&	{\bf  AM 1754-634 NED03}$^{\star}$	&	166	&				&		\\
10	&	(48.8, $-$28.7)	&	CGCG 374-029	&	183	&	PC 2055+0126	&	125	\\
		&			&				&		&	{\bf  Mrk 510}		&	152	\\
19	&	(63.5, $-$40.2)	&	PC 2207+0122$^{\star}$	&	98	&	PKS~2201+044	&	109	\\
		&			&				&		&	{\bf  NGC 7189}$^{\star}$	&	151 \\
7	&	(58.8, $-$42.4)	&	PC 2207+0122$^{\star}$	&	189	&	{\bf  NGC 7189}$^{\star}$	&	108 \\
16	&	($-$170.6, $-$45.7)	&	NGC 1358$^{\star}$	&	51	&	Mrk 612		&	81 \\
	&				&	NGC 1346$^{\star}$	&	66	&	Mrk 609$^{\star}$	&	133 \\
		&			&	NGC 1320		&	130	&	{\bf  KUG 0322-063A}$^{\star}$	&	137 \\
12	&	($-$165.9, $-$46.9)	&	NGC 1346$^{\star}$	&	152	&	Mrk 609$^{\star}$		&	169 \\
		&			&	NGC 1358$^{\star}$	&	166	&	{\bf  KUG 0322-063A}$^{\star}$	&	171 \\
15	&	(88.8, $-$47.1)	&	{\bf  NGC 7626}		&	86	&	{\bf  2MASX J23274259+0845298}	&	122 \\
		&			&	NGC 7591		&	186	&	NGC 7674		&	125 \\
24	&	(12.1, $-$49.0)	&	NGC 7130		&	117	&	6dF J2132022-334254	&	127 \\
		&			&	NGC 7135		&	126	&					&	\\
22	&	($-$163.8, $-$54.4)	&	NGC 1204		&	96	&	MCG-02-08-039	&	174	\\
9	&	(4.2, $-$54.9)	&	IC 5169		&	131	&	ESO 404-IG042	&	76	\\
		&			&				&		&	PKS~2158-380	&	126 \\
6	&	($-$75.6, $-$78.6)	&	NGC 0424		&	25	&	ESO 351-G025	&	153 \\
		&			&				&		&	{\bf ESO 352-G048}	&	163 
\enddata
\tablecomments{$\theta$ denotes the separation between an UHECR event and an AGN. 
Stars denote AGN falling within the $3.2^{\circ}$ search radius of two different events. 
AGN marked with bold font are those which \emph{do not} appear 
in the \citet{VC2006} catalog.}
\end{deluxetable*}

\begin{deluxetable*}{lllllllllllll}
\vspace*{-7mm}
\ifthenelse{\equal{\ms}{preprint}}%**********************************************
{\tabletypesize{\scriptsize}}
{
\tabletypesize{\footnotesize}
\rotate
}
\tablecaption{\label{t2} Properties of selected AGN}
\tablecolumns{14}
\tablewidth{0pt}
\tablehead{Name & Type & RA & DEC & $z$ & $d_{\rm L}$ & UHECR & $\theta$ & $S_{\rm R}$ & Ref. & $S_{\rm FIR}$ & $S_{\rm X}$ & Ref.\\
 & & (J2000.0) & (J2000.0) & & [Mpc] & $(\ell\,[^{\circ}], \, b\,[^{\circ}])$ & [$\arcmin$] & [mJy] & & [Jy] & [cgs] & \\
(1) & (2) & (3) & (4) & (5) & (6) & (7) & (8) & (9) & (10) & (11) & (12) & (13)}
\startdata
NGC 5506 & Sy2 & 14h13m14.8s & $-$03d12m27s & 0.0062 & 29 & ($-$21.8, 54.1) & 38 & 227 & G06 & 8.4 & 12 & S06 \\
UM 653 & Sy2 & 14h16m15.5s & $-$01d27m53s & 0.0365 & 158 & ($-$21.8, 54.1) & 136 & 0.9$^c$ & B95 & 0.4 & --- & ---\\
UM 654 & Sy2 & 14h16m19.7s & $-$01d25m18s & 0.0369 & 160 & ($-$21.8, 54.1) & 138 & $<$0.2$^c$ & B95 & --- & --- & ---\\
UM 625 & Sy2 & 14h00m40.6s & $-$01d55m18s & 0.0250 & 109 & ($-$21.8, 54.1) & 171 & 1.1$^c$ & B95 & 0.3 & --- & --- \\
ESO 575-IG016 & S2 & 12h52m36.2s & $-$21d54m46s & 0.0229 & 100 & ($-$57.2, 41.8) & 55 & 1.7$^c$ & C98 & --- & --- & --- \\ 
MCG-03-32-017 & L(?) & 12h38m00.5s & $-$20d07m51s & 0.0280 & 123 & ($-$57.2, 41.8) & 191 & $<$0.8$^c$ & C98 & --- & --- & ---\\
TOLOLO 20 & Sy1 & 12h12m20.0s & $-$28d48m46s & 0.0300 & 131 & ($-$65.1, 34.5) & 133 & $<$0.5$^c$ & C98 & --- & --- & ---\\
2MASX J1323 & L(?) & 13h23m02.4s & $-$34d52m47s & 0.0261 & 113 & ($-$50.8, 27.6) & 55 & 10.8$^c$ & C98 & --- & --- & ---\\
TOLOLO 81 & Sy2 & 13h19m38.6s & $-$33d22m54s & 0.0291 & 127 & ($-$50.8, 27.6) & 95 & $<$0.8$^c$ & C98 & --- & --- & ---\\
NGC 5140 & L(?) & 13h26m21.7s & $-$33d52m06s & 0.0129 & 58 & ($-$50.8, 27.6) & 116 & 29.2 & S89 & 0.8 & --- & --- \\
ESO 444-G018 & L(?) & 13h22m56.8s & $-$32d43m42s & 0.0292 & 127 & ($-$50.8, 27.6) & 144 & 7.5$^c$ & C98 & --- & --- & ---\\
NGC 2907 & L(?) & 09h31m36.7s & $-$16d44m05s & 0.0070 & 33 & ($-$109.4, 23.8) & 94 & 4.4$^c$ & M92 & 0.4 & --- & --- \\
NGC 2989 & L(?) & 09h45m25.2s & $-$18d22m26s & 0.0138 & 62 & ($-$109.4, 23.8) & 182 & 7.6$^c$ & C98 & 1.7 & --- & ---\\
NGC 5128 (Cen~A) & FR~I & 13h25m27.6s & $-$43d01m09s & 0.0018 & 3.4 & ($-$51.4, 19.2) & 54 & 53792 & G94 & 162 & 30 & B06 \\ 
 & & & & & & ($-$51.2, 17.2) & 139 & & & & & \\ 
NGC 5064 & L & 13h18m59.9s & $-$47d54m31s & 0.0099 & 45 & ($-$51.2, 17.2) & 162 & 4.0$^c$ & M03 & 3.3 & --- & --- \\
 & & & & & & ($-$52.8, 14.1) & 47 & & & & & \\ 
NGC 5244 & L(?) & 13h38m41.7s & $-$45d51m21s & 0.0085 & 38 & ($-$51.2, 17.2) & 164 & 5.7$^c$ & M03 & 1.9 & --- & --- \\
IRAS 13028-49 & Sy(?) & 13h05m45.5s & $-$49d25m22s & 0.0012 & 8.3 & ($-$52.8, 14.1) & 118 & $<$22$^c$ & B99 & 6.1 & --- & ---\\
NGC 4945 & Sy(?) & 13h05m27.5s & $-$49d28m06s & 0.0019 & 3.8 & ($-$52.8, 14.1) & 121 & 2953 & G94 & 359 & 0.5 & L04 \\
IC 4518A & S2 & 14h57m41.2s & $-$43d07m56s & 0.0163 & 70 & ($-$34.4, 13.0) & 66 & 64.9$^c$ & M03 & --- & 1.9$^h$ & B07 \\
1RXS J174155 & Sy1 & 17h41m55.3s & $-$12d11m57s & 0.0370 & 156 & (15.4, 8.4) & 106 & 1.4$^c$ & C98 & --- & 2.9$^h$ & B07\\
HB91 1739-126 & Sy1 & 17h41m48.7s & $-$12d41m01s & 0.0370 & 156 & (15.4, 8.4) & 122 & $<$0.5$^c$ & C98 & --- & --- & ---\\
WKK 2031 & S2 & 13h15m06.3s & $-$55d09m23s & 0.0308 & 133 & ($-$52.3, 7.3) & 83 & 154 & G94 & 41.0 & --- & --- \\
WKK 4374 & Sy2 & 14h51m33.1s & $-$55d40m38s & 0.0180 & 77 & ($-$41.7, 5.9) & 167 & 6.2$^c$ & B99 & --- & 2.0$^h$ & B07\\
DZOA 4653-11 & Sy1 & 13h47m36.0s & $-$60d37m04s & 0.0129 & 56 & ($-$49.6, 1.7) & 40 & $<$3.8$^c$ & B99 & --- & 7.8$^h$ & B07\\
PKS~1343-60 (Cen~B) & FR~I & 13h46m49.0s & $-$60d24m29s & 0.0129 & 56 & ($-$49.6, 1.7) & 41 & 27100 & W90 & --- & 0.2$^l$ & M05 \\ 
ESO 139-G012 & Sy2 & 17h37m39.1s & $-$59d56m27s & 0.0170 & 71 & ($-$27.6, $-$16.6) & 109 & 5.2$^c$ & M03 & 0.7 & --- & --- \\
 & & & & & & ($-$27.7, $-$17.0) & 139 & & & & & \\ 
AM 1754-634 & Sy(?) & 18h00m10.9s & $-$63d43m34s & 0.0157 & 65 & ($-$27.6, $-$16.5) & 191 & $<$0.9$^c$ & B99 & --- & --- & ---\\
 & & & & & & ($-$27.7, $-$17.0) & 166 & & & & & \\
PC 2055+0126 & S(?) & 20h58m18.2s & $+$01d38m00s & 0.0260 & 105 & (48.8, $-$28.7) & 125 & $<$0.8$^c$ & C98 & --- & --- & --- \\
Mrk 510 & Sy(?) & 21h09m23.0s & $-$01d50m17s & 0.0195 & 77 & (48.8, $-$28.7) & 152 & 48.4$^c$ & W92 & --- & --- & --- \\
CGCG 374-029 & S1 & 20h55m22.3s & $+$02d21m16s & 0.0136 & 52 & (48.8, $-$28.7) & 183 & 1.8$^c$ & C98 & --- & --- & --- \\
PC 2207+0122 & S(?) & 22h10m30.0s & $+$01d37m10s & 0.0130 & 49 & (63.5, $-$40.2) & 98 & $<$0.2$^c$ & B95 & --- & --- & --- \\
 & & & & & & (58.8, $-$42.4) & 189 & & & & & \\
PKS~2201+044 & BL & 22h04m17.6s & $+$04d40m02s & 0.0270 & 108 & (63.5, $-$40.2) & 109 & 530 & W90 & --- & 0.2$^l$ & B94 \\
NGC 7189 & L & 22h03m16.0s & $+$00d34m16s & 0.0302 & 122 & (63.5, $-$40.2) & 151 & 13.0$^c$ & C02 & 3.1 & --- & --- \\
 & & & & & & (58.8, $-$42.4) & 108 & & & & & \\ 
NGC 1358 & Sy2 & 03h33m39.7s & $-$05d05m22s & 0.0134 & 54 & ($-$170.6, $-$45.7) & 51 & 8.7$^c$ & C98 & 0.4 & 0.04 & U05 \\
 & & & & & & ($-$165.9, $-$46.9) & 166 & & & & & \\ 
Mrk 612 & Sy2 & 03h30m40.9s & $-$03d08m16s & 0.0205 & 83 & ($-$170.6, $-$45.7) & 81 & 5.1$^c$ & C98 & 1.2 & 0.04 & G05 \\
NGC 1320 & Sy2 & 03h24m48.7s & $-$03d02m32s & 0.0089 & 35 & ($-$170.6, $-$45.7) & 130 & 3.3 & G06 & 2.2 & --- & --- \\
Mrk 609 & Sy2 & 03h25m25.3s & $-$06d08m38s & 0.0345 & 143 & ($-$170.6, $-$45.7) & 133 & 12.5$^c$ & C98 & 2.6 & --- & --- \\
 & & & & & & ($-$165.9, $-$46.9) & 169 & & & & & \\ 
KUG 0322-063 & Sy1 & 03h25m11.6s & $-$06d10m51s & 0.0338 & 140 & ($-$170.6, $-$45.7) & 137 & 11.3$^c$ & C98 & 2.1 & --- & --- \\
 & & & & & & ($-$165.9, $-$46.9) & 171 & & & & & \\ 
NGC 1346 & L(?) & 03h30m13.3s & $-$05d32m36s & 0.0135 & 54 & ($-$170.6, $-$45.7) & 66 & 10.4$^c$ & C98 & 3.1 & --- & --- \\
 & & & & & & ($-$165.9, $-$46.9) & 152 & & & & & \\
NGC 7626 & L & 23h20m42.5s & $+$08d13m01s & 0.0114 & 42 & (88.8, $-$47.1) & 86 & 210 & W90 & --- & 0.03$^l$ & T05 \\
2MASX J2327 & Sy2 & 23h27m42.6s & +08d45m30s & 0.0294 & 118 & (88.8, $-$47.1) & 122 & $<$0.5$^c$ & C98 & --- & --- & ---\\
NGC 7674 & Sy2 & 23h27m56.7s & $+$08d46m45s & 0.0289 & 116 & (88.8, $-$47.1) & 125 & 90.6$^c$ & C02 & 5.6 & 0.05 & L04 \\
NGC 7591 & L & 23h18m16.3s & $+$06d35m09s & 0.0165 & 64 & (88.8, $-$47.1) & 186 & 21.4$^c$ & C02 & 7.2 & --- & --- \\
NGC 7130 & Sy2 & 21h48m19.5s & $-$34d57m05s & 0.0161 & 64 & (12.1, $-$49.0) & 117 & 63.3$^c$ & M03 & 16 & 0.006 & Gm06 \\
NGC 7135 & L(?) & 21h49m46.0s & $-$34d52m35s & 0.0088 & 33 & (12.1, $-$49.0) & 126 & 2.3$^c$ & C98 & 0.2 & --- & --- \\ 
6dF J2132022 & Sy1 & 21h32m02.2s & $-$33d42m54s & 0.0293 & 120 & (12.1, $-$49.0) & 127 & 1.1$^c$ & C98 & --- & --- & ---\\
NGC 1204 & L & 03h04m39.9s & $-$12d20m29s & 0.0143 & 57 & ($-$163.8, $-$54.4) & 96 & 9.4$^c$ & C90 & 7.8 & --- & --- \\
MCG-02-08-039 & Sy2 & 03h00m30.6s & $-$11d24m57s & 0.0299 & 123 & ($-$163.8, $-$54.4) & 174 & 3.6$^c$ & C98 & 0.5 & --- & --- \\
ESO 404-IG042 & S2 & 22h13m17.5s & $-$37d00m58s & 0.0340 & 140 & (4.2, $-$54.9) & 76 & 2.2$^c$ & C98 & 0.7 & --- & --- \\
PKS~2158-380 & FR~II & 22h01m17.1s & $-$37d46m24s & 0.0333 & 137 & (4.2, $-$54.9) & 126 & 590 & W90 & 0.3 & --- & --- \\
IC 5169 & Sy2 & 22h10m10.0s & $-$36d05m19s & 0.0104 & 39 & (4.2, $-$54.9) & 131 & 8.3$^c$ & M03 & 3.4 & --- & --- \\
NGC 0424 & Sy1 & 01h11m27.6s & $-$38d05m00s & 0.0118 & 46 & ($-$75.6, $-$78.6) & 25 & 9.6$^c$ & C98 & 1.8 & 0.1 & U05 \\
ESO 351-G025 & Sy2 & 00h58m22.3s & $-$36d39m37s & 0.0346 & 143 & ($-$75.6, $-$78.6) & 153 & 2.4$^c$ & C98 & --- & --- & --- \\
ESO 352-G048 & Sy2 & 01h20m54.7s & $-$36d19m26s & 0.0322 & 132 & ($-$75.6, $-$78.6) & 163 & $<$0.5$^c$ & C98 & --- & --- & ---
\enddata
\tablecomments{\scriptsize
[1] Name of the source. 
[2] AGN classification (Sy1/Sy2: Seyfert galaxy of the type~1/type~2; L: LINER; 
FR~I: radio galaxy of the FR~I type; BL: BL Lacertae object). 
[3-4] Equatorial coordiantes (J2000.0). 
[5] Redshift of the source. 
[6] Luminosity distance for the assumed cosmology 
($H_0 = 73$ km s$^{-1}$ Mpc$^{-1}$, $\Omega_{\rm M} = 0.27$, $\Omega_{\Lambda} = 0.73$)
except for NGC 5128 (Cen~A) and NGC 4945, where the distances are known
\citep{2007AJ....133..504K}.
[7] Galactic coordiantes for the nearby UHECR event. 
[8] Separation between an AGN and a nearby UHECR event. 
[9] The total $5$\,GHz flux in mJy units ($^c$fluxes converted from the ones provided at lower 
frequencies assuming radio spectral index $\alpha = 0.7$). 
[10] References for column~9. 
[11] The total $60$ $\mu$m flux in Jy units from the IRAS survey \citep{Moshir1990}. 
[12] The total observed $2-10$ keV flux in $\times 10^{-11}$ erg cm$^{-2}$ s$^{-1}$ units 
($^l$fluxes provided at lower photon energy range: $0.5-7$ keV in \citet{Marshall2005}, 
or $0.1-2.0$ keV in \citet{Brinkmann1994} and \citet{Tajer2005}; $^h$fluxes converted from the 
ones provided at higher photon energy range $40-100$\,keV in \citet{Bird2007}, assuming X-ray 
photon index $\Gamma_X = 2.0$). 
[13] References for column~12.
}
\tablerefs{\scriptsize
(B95) \citealt{Becker1995};
(B06) \citealt{Beckmann2006};
(B07) \citealt{Bird2007};
(B99) \citealt{Bock1999};
(B94) \citealt{Brinkmann1994};
(C90) \citealt{Condon1990};
(C98) \citealt{Condon1998};
(C02) \citealt{con02};
(G06) \citealt{gal06};
(Gm06) \citealt{Gonzalez-Martin2006};
(G94) \citealt{Gregory1994};
(G05) \citealt{Guainazzi2005};
(L04) \citealt{Lutz2004};
(M05) \citealt{Marshall2005};
(M03) \citealt{Mauch2003};
(M92) \citealt{Mollenhoff1992};
(S89) \citealt{Sadler1989};
(S06) \citealt{Shinozaki2006};
(T05) \citealt{Tajer2005};
(U05) \citealt{Ueda2005};
(W92) \citealt{WB1992};
(W90) \citealt{WO1990}.
}
\end{deluxetable*}

We collected AGN which are possibly associated with the UHECR events 
by searching \emph{both} the NASA Extragalactic Database
(NED)\footnote{\texttt{http://nedwww.ipac.caltech.edu/}} and the \citet{VC2006}
catalog. 
We also extend the search out to redshifts $z \leq 0.037$ ($\leq$150 Mpc) 
which doubles the distance
at which maximum significance was found by the \AC{} ($z \leq 0.018$,
$\leq$75 Mpc); this effectively covers the range of horizons for super-GZK particles
\citep{Harari2006,Abraham2008a}.

Table~\ref{t1} lists the AGN found within the $3.2^{\circ}$ search radius 
around each UHECR event reported by \PAO{}
\citep{Abraham2008a}, 
divided into two groups: those within redshift 
range $z\le 0.018$ and $0.018<z\le 0.037$ (corresponding approximately to the 
luminosity distances $d_{\rm L} \leq 75$ Mpc and
$75$\,Mpc\,$\leq d_{\rm L} \leq 150$ Mpc, respectively, Table~\ref{t2}).
In total, we have selected 54 active galaxies, 27 per each redshift bin. 
Note that
six UHECR events lack any selected AGN counterpart located within 
$75$ Mpc, and for two
of them we also did not find any possible AGN association up to a distance 
$150$ Mpc.
Four of these are located at low Galactic latitudes, $|b| < 12^{\circ}$.
Obviously, most of the events possess multiple AGN ``counterparts'' within the 
assumed search radius and redshift range.

\begin{deluxetable}{lllll}
\tabletypesize{\footnotesize}
\tablecaption{\label{t3} SDSS AGN \citep{Hao2005}.}
\tablecolumns{5}
\tablewidth{0pt}
\tablehead{Event Number & Name & Type & $z$ & $\theta$\\
 & & & & [$\arcmin$]\\
(1) & (2) & (3) & (4) & (5)}
\startdata
7	&	SDSS J215259.07-000903.4	&	Sy2	&	0.028	&	188\\
	&	SDSS J220515.43-010733.3	&	Sy2	&	0.032	&	7\\
12	&	SDSS J033458.00-054853.2	&	Sy1	&	0.018	&	123\\
 &       SDSS J034545.16-071526.8     &   Sy2   &  0.022  &  172\\
 &       SDSS J033955.97-063228.9     &   Sy2   &  0.031  &  113\\
 &       SDSS J033713.31-071718.0     &   Sy2   &  0.033  &  53\\
 &       SDSS J032329.63-062944.1     &   Sy2   &  0.034  &  182\\
 &       SDSS J034330.25-073507.4     &   Sy2   &  0.036  &  135\\
16	&	SDSS J033458.00-054853.2	&	Sy1	&	0.018	&	97\\
 &       SDSS J033955.97-063228.9     &   Sy2   &  0.031  &  179\\
 &       SDSS J033713.31-071718.0     &   Sy2   &  0.033  &  190\\
 &       SDSS J032329.63-062944.1     &   Sy2   &  0.034  &  168
\enddata
\tablecomments{
[1] UHECR Event Number.
[2] Name of the source (note that source positions are implicit in the SDSS names).
[3] AGN classification (Sy1/Sy2: Seyfert galaxy of the type~1/type~2).
[4] Redshift of the source. 
[5] Separation between an AGN and a nearby UHECR event.
}
\end{deluxetable}

We emphasize that the \citet{VC2006} and NED assemblages are not
complete AGN catalogs. 
Additionally, we have found several inconsistencies between these two
databases regarding source classifications.
This results in a few sources showing some
(weak) level of AGN activity (according to one but not the other catalog) 
which might fulfill the criteria 
for a possible association with the detected UHECR events,  
but these are not included in our dataset\footnote{
For example, 
we did not include the object ESO 383-G18 in our dataset, since its more accurate position from 
2MASS as given in NED makes it $192.2\arcmin$ away from the location of CR\#2. 
With the less accurate position listed by \citet{VC2006}, however, this source 
is $<$$192\arcmin$ from the considered CR event.
}. 
However, a few other objects which do not obviously possess
an active nucleus (but rather only \hii{} central activity, like NGC~4945,
NGC~5244, NGC~2989, NGC~7135, and especially nearby IRAS 13028-49\footnote{See
\citet{str92}.}) are included and classified as ``possibly LINERs.''
Furthermore, the regions around 3 events (\#7, \#12, \#16) are covered in the
Sloan Digital Sky Survey \citep[SDSS;][]{York2000} leading to a somewhat misleading
higher density of known AGN in these regions in comparison to other event regions.
We have therefore omitted the 12 SDSS AGN in the \citet{VC2006} list falling within
the search volume from Tables~\ref{t1}, \ref{t2} and listed them separately 
in Table~\ref{t3}. 
These objects, all classified as Seyferts, are
all
located
at redshifts $z > 0.018$ and display similar bulk properties to the Seyferts
we have included in our discussion.
Most of the selected objects (50 out of 54) are relatively
weak Seyfert galaxies and LINERs, while only a small fraction (4) are bright
and well established jet sources (three radio galaxies and one BL Lac object).

It is probable that more Seyferts/LINERs are missing (as evident in the SDSS
covered regions) and/or mis-classified in the presented AGN collection.
However, we expect much fewer radio galaxies/BL Lacs to be omitted. For an
order of magnitude estimate, we derive 
an average number density 
of radio galaxies within the redshift range $z \leq 0.037$ of $\sim 3 \times
10^{-6}$ Mpc$^{-3}$ (consistent with an estimate of the space density of 
UHECR sources given below)
assuming the total radio luminosity range between
$10^{38}$ erg s$^{-1}$ and $\geq 10^{42}$ erg s$^{-1}$. 
This was derived
using the $151$ MHz luminosity function of low-power radio sources (including
classical FR~I, and FR~II radio galaxies with weak/absent emission lines) of
\citet{wil01}, converting their model C to the modern cosmology adopted in
this paper \citep[discussed in detail in][]{sta06} and the total 151 MHz
luminosities to 5 GHz ones assuming a radio spectral index of $\alpha_R =
0.7$. The derived number density of local radio galaxies is therefore about
three orders of magnitude lower than the number density of Seyfert galaxies
(see below), resulting in a total of $\sim$42 expected radio galaxies within
the co-moving volume of the local Universe $V = 0.014$ Gpc$^3$. We note that
with the assumed minimum radio luminosity one order of magnitude lower, i.e.
$10^{37}$ erg s$^{-1}$, the number density of radio galaxies increases by a
factor of $\sim$4. At such low radio luminosities, however, the luminosity
function constructed by \citet{wil01} is not well determined.

With the small number of nearby radio galaxies, we expect that the majority of
these objects have already been identified as such through radio imaging
studies. The luminosities of the radio galaxies/BL Lac in our sample (down to
$4\times10^{39}$ erg s$^{-1}$, the 5 GHz luminosity of Cen~A), correspond to a relatively bright
flux limit ($\sim$100 mJy source\footnote{For $\alpha_{\rm R}=0.7$, this
corresponds to $\sim$250 mJy at 1.4 GHz.} at $D=160$ Mpc).  The morphologies of
such bright radio sources have been well surveyed \citep[e.g.,][and references
therein]{Sadler1989,1995AJ....109...14O} although admittedly, there is sparser coverage
in some parts of the sky, in, e.g., the Southern hemisphere
\citep{2006AJ....131..114B}. 
The Galactic plane is particularly susceptible to having
unidentified local radio galaxies but the radio imaging surveys toward the
plane are maturing \citep{2006AJ....131.2525H} and hard X-ray surveys are beginning to
help rectify the situation \citep[e.g.,][]{2007MNRAS.382..937M}.

\placetable{Table~\ref{t2}}
\placetable{Table~\ref{t3}}

Table~\ref{t2} lists the main properties of all the selected AGN collected
from the literature, in particular the total radio fluxes at $5$\,GHz 
(or upper 
limits for these), {\it IRAS} fluxes at $60$\,$\mu$m (if available), 
and
$2-10$\,keV fluxes (again, if available). 
In some cases, the flux conversion from values provided at lower radio 
frequencies (or in higher X-ray photon energy range)
was necessary, and we performed this assuming typical values of the radio 
spectral
index $\alpha_R = 0.7$ (X-ray photon index $\Gamma_X = 2.0$). 
Note that 12 sources listed in Table~\ref{t2} have only upper limits
for the total radio emission
which were derived by us from the {\it FIRST} \citep{Becker1995}, {\it
NVSS} \citep{Condon1998}, or {\it SUMMS} \citep{Bock1999} surveys, and
most of them are indeed very weak radio sources ($<$mJy level).

Only in the case of the four selected FR~I/BL
Lac sources 
(Cen~A, Cen~B, PKS~2158--380, PKS~2201+044)
is the presence of relativistic jets -- as ones considered
in different models for the acceleration of UHECRs
-- certain. 
As for the rest of the selected
AGN, one should ask if jet activity can be ascertained
and, if so, what are the properties of the jets in these sources. 
Are
these jets relativistic? What are their inclinations? Do they differ
somehow from the jets observed in other Seyfert and LINER galaxies?
Such questions are relevant, because the space density of the local
low-luminosity AGN, just like the ones selected in our sample, is
high. 
For example, the space density of bright galaxies ($-22\ {\rm mag} \leq
M_{B} \leq -18$ mag) showing Seyfert activity is $\sim$$1.25 \times
10^{-3}$ Mpc$^{-3}$ \citep{ulv01}\footnote{This is about $\sim$$10^3$
times larger than the local space density of bright quasars, and $\sim$$10$ 
times smaller than the space density of ``regular'' galaxies
with comparable brightness \citep{ulv01}.} 
which is $\sim$$10-10^3$ times larger than the space density of UHECR sources, 
$\sim$$10^{-6}-10^{-4}$ Mpc$^{-3}$, derived from a comparison of simulation of 
particle propagation in the local universe and the Akeno Giant Air Shower Array (AGASA) data 
\citep{Blasi2004,Sigl2004,Takami2006}. 
Note that \citet{Dubovsky2000} gives an estimate of the space density of UHECR sources
$\sim$$6\times10^{-3}$ Mpc$^{-3}$ based on 7 observed doublets (only two events have 
estimated energy $>$100 EeV) and assuming the distance
$\la$25 Mpc for protons $>$100 EeV. When rescaled to the distance $\la$75 Mpc, 
it gives $\sim$$2\times10^{-4}$ Mpc$^{-3}$, consistent with other estimates.
Recently \citet{Takami2008} estimated a space density of UHECR sources $\sim$$10^{-4}$ Mpc$^{-3}$ based on the \PAO{} data.
Thus, restricting the
investigations to redshifts $z \leq 0.037$, corresponding to the
luminosity distance $d_{\rm L} \leq 156.4$ Mpc and co-moving
volume $V = 0.014$ Gpc$^3$, 
there are 
$\sim$$1.8\times10^4$ low-luminosity AGN with surface density (if distributed
isotropically over the whole sky)
$\sim$$1.4\times10^3$ per steradian.
In other words, there should be 
$\sim$14 AGN
within the search radius $3.2^\circ$.
The much lower rate of our
possible identification given in Table~\ref{t1} is due to
incompleteness/lack of the AGN surveys in different parts of the sky,
and especially within the Galactic Plane (see a decrease in the
identification rate for Galactic latitudes $|b| < 12^{\circ}$ in
Table~\ref{t1}). 
Therefore, if the selected Seyferts and LINERs do not differ from
the other local AGN of the same types, the claimed correlation
between the UHECR events and local active galaxies should be
considered as rather unlikely, resulting from a chance coincidence, 
if the production of the highest 
energy CRs is not episodic in nature, but operates in a single object 
on long ($\geq$ Myr) timescales. 
In
addition, if the selected sources do not show significant jet
activity, there are no reasons for expecting them to accelerate CRs
up to 10--100 EeV energies.

\begin{figure}
\begin{center}
\includegraphics[height=3.5in,angle=270]{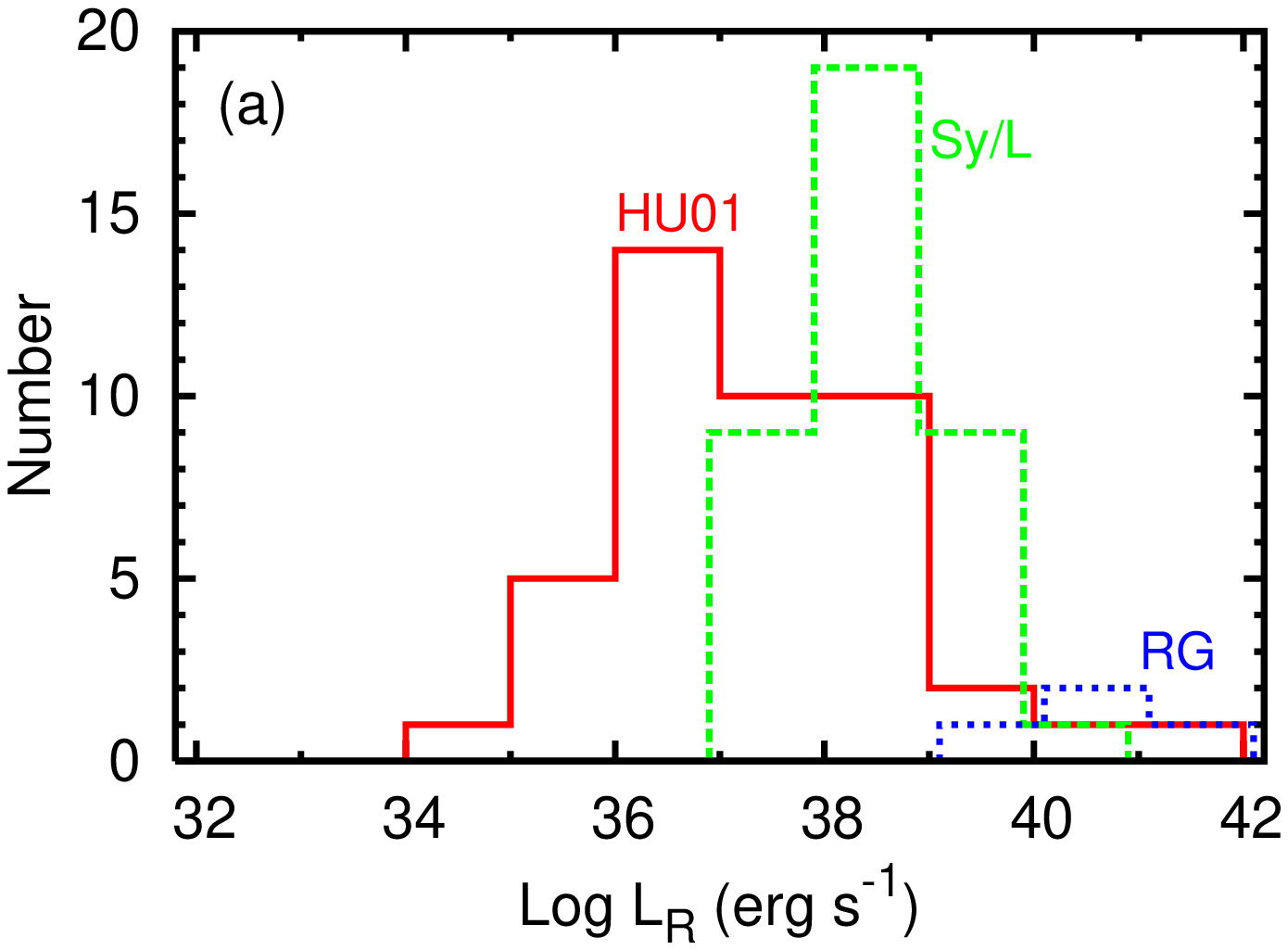}
\includegraphics[height=3.5in,angle=270]{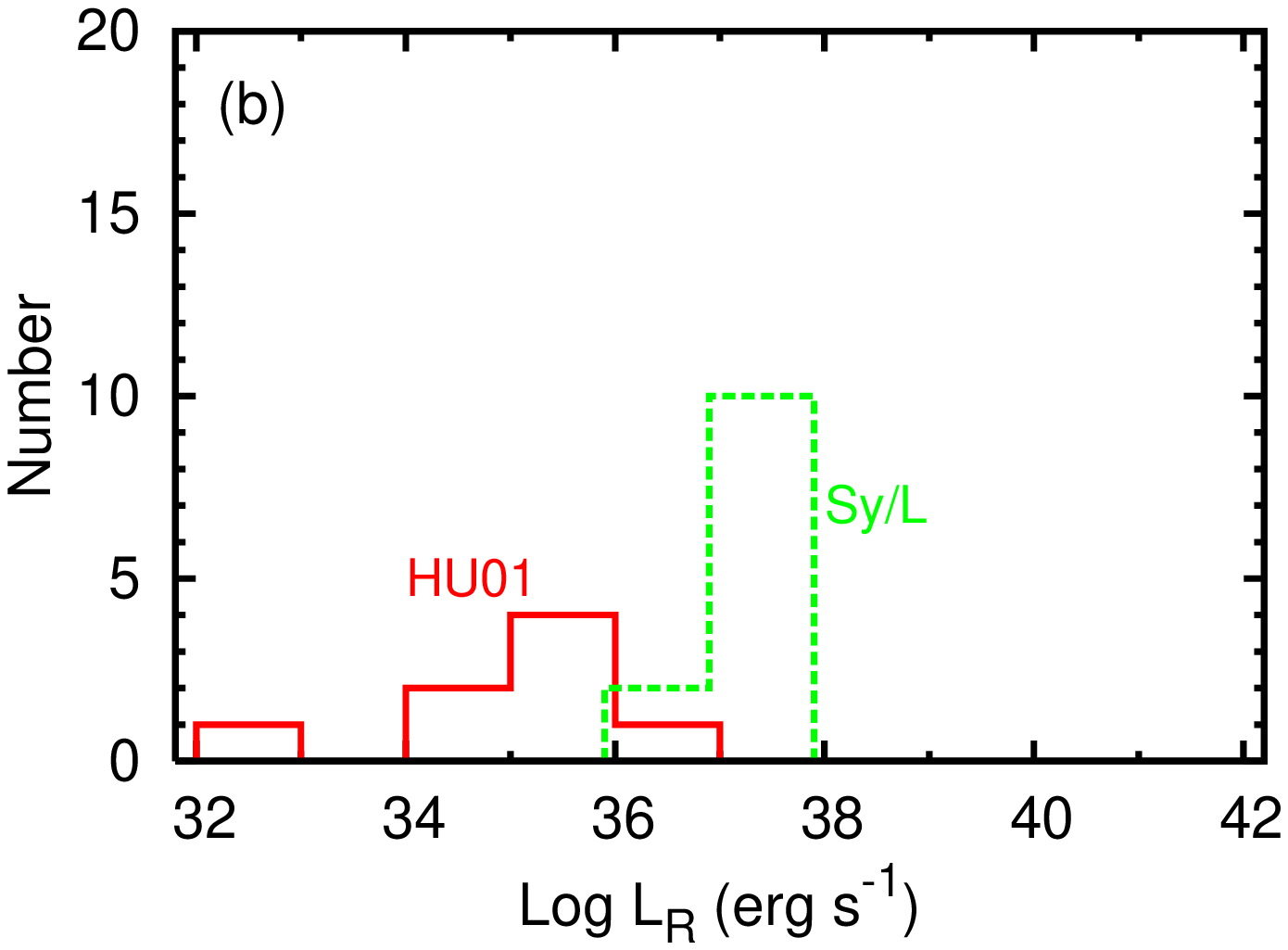}
\end{center}
\caption{{\it (a)} Number distributions of 5~GHz total radio powers
for Seyfert galaxies detected at radio frequencies from the sample 
constructed by \citet[][red solid]{ho01}, Seyfert/LINER galaxies 
from our sample detected at radio frequencies (green dashed), and radio 
galaxies from our sample (blue dotted). 
{\it (b)} Number distributions of upper limits to 5~GHz total radio powers
for Seyfert galaxies detected at radio frequencies from the sample 
constructed by \citet[][red solid]{ho01}, and Seyfert/LINER galaxies 
from our sample detected at radio frequencies (green dashed).
The histograms are slightly shifted relative to each other in horizontal direction for clarity.
}
\label{fig:hist}
\end{figure}

\placefigure{figure~\ref{fig:hist}}

A brief summary of the radio properties of local
low-luminosity AGN given in \S\ref{agn_ph} 
allows us to conclude that the particular Seyferts
and LINERs selected in Table~\ref{t2} are most likely jetted, but do not
differ from the other analogous sources of the same type. 
In particular, complex radio morphologies consisting of compact cores,
one-sided jets, and extended halos, as found in several of the
selected objects, are typical for the Seyfert/LINER-type jet
activity (see \S\ref{agn_ph}).
Such jets are expected to be
sub-relativistic ($v<0.25 c$ on pc-scales), low-power ($L_{\rm j}
\leq 10^{43}$ erg s$^{-1}$), precessing and short-lived ($t_{\rm j}
\lesssim 10^5$ yrs). 
Indeed, 5 GHz powers for the selected
Seyferts and LINERs, being in a range $L_{\rm 5\,GHz} \sim
10^{37}-10^{42}$ erg s$^{-1}$, are typical for the other AGN of the
same kind, and are significantly lower than the radio powers of 4
selected radio galaxies/BL Lac. 
The median values of these, $\sim$$2
\times 10^{38}$ erg s$^{-1}$ (including only radio-detected sources), 
seems to be higher than the
appropriate median values given by \citet{ho08}, but this may be
simply due to selection effects. 
In fact, all 12 
objects for which only upper limit regarding the radio fluxes are provided,
have $L_{\rm 5\,GHz} < 5 \times 10^{37}$ erg s$^{-1}$. 

Figure~\ref{fig:hist}a shows the number distributions of 5~GHz total
radio powers for Seyfert galaxies detected at radio frequencies from
the optically selected sample constructed by \citet{ho01}, the
Seyfert/LINER galaxies
from our sample detected at radio frequencies, and radio
galaxies from our sample. At first glance, our Seyferts/LINERs appear
systematically more radio luminous in comparison to the \citet{ho01}
sample by 1--2 dex.  However, we note that the upper limits we obtain from the
available ``all-sky" maps (NVSS, SUMMS) are systematically higher than
the ones obtained by \citeauthor{ho01} from their pointed observations
(Figure~\ref{fig:hist}b) also by 1--2 orders of magnitude, making it
difficult to compare the two distributions directly.
The radio galaxies are characterized by larger radio luminosities than
Seyferts/LINERs, as expected.

The range of X-ray luminosities (if available) of objects in our sample is 
in the range $L_{\rm 2-10\,keV} \sim
10^{40}-10^{44}$ erg s$^{-1}$ \citep[with the median $\sim$$3 \times
10^{41}$ erg s$^{-1}$ again slightly higher than the one provided
by][]{ho08}, which is comparable to the typical 2--10 keV luminosities
observed in local low-power AGN \citep{pan07}. 
Moreover, the
logarithm of the ratio of the X-ray and radio luminosities,
$\log(L_{\rm 5\,GHz}/L_{\rm 2-10\,keV})$, although widely scattered in
a range between ($-1.3$) and ($-5.6$) with median ($-3.2$), 
is in agreement with the values
found in other local low-luminosity AGN \citep[see][]{pan07}.

\begin{figure}
%\centerline{
%\includegraphics[width=3.5in,angle=270]{f2.ps}
\includegraphics[height=3.6in,angle=270]{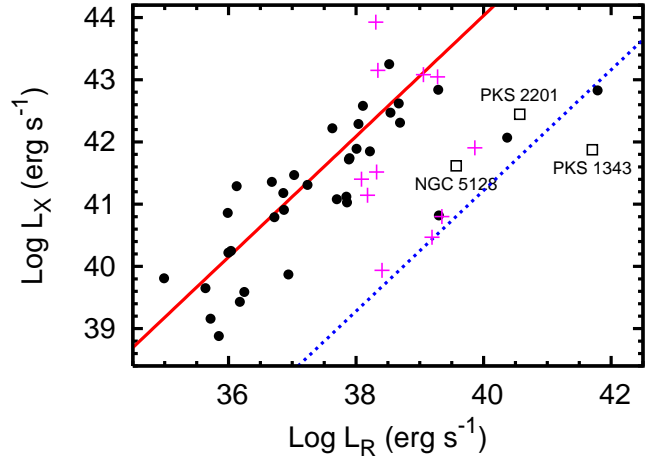}
%}
\caption{X-ray ($2-10$~keV) luminosity versus $5$~GHz total luminosity for
Seyfert galaxies from the sample constructed by 
\citet[][{\it black circles}]{pan07}, 
Seyfert/LINER galaxies from our sample with the 
X-ray and $5$~GHz fluxes provided ({\it magenta crosses}), 
radio galaxies from our sample with the X-ray fluxes provided 
({\it open squares}). 
In the case of the sample of \citet{pan07}, only sources detected at X-ray and
radio frequencies were considered. Red solid line indicates the best fit $\log L_X
=0.97 \, \log L_R + 5.23$ for Seyfert galaxies, and blue dotted line denotes the best fit 
$\log L_X =0.97 \, \log L_R + 2.42$ for low-luminosity radio galaxies, both as given by
\citeauthor{pan07}
} 
\label{fig:L1}
\end{figure}

\placefigure{figure~\ref{fig:L1}}

Figure~\ref{fig:L1} shows the X-ray ($2-10$~keV) luminosity versus 5~GHz total luminosity for
Seyfert galaxies from the sample constructed by 
\citet{pan07}, Seyfert/LINER galaxies from our sample with the 
X-ray and 5~GHz fluxes provided, and 
radio galaxies from our sample with the X-ray fluxes provided.
In the case of the sample of \citet{pan07}, only sources detected at X-ray and
radio frequencies were considered. 
The bulk of our selected Seyferts/LINERs match well the 
$L_X - L_R$ correlation established for nearby Seyferts. Our selected radio galaxies and
a few Seyferts are over-luminous in the radio for a given X-ray luminosity when compared
to the Seyferts. 
However, they match well the $L_X - L_R$ correlation established for 
nearby ``low-luminosity radio galaxies'' \citep[LLRGs,][]{pan07}.

Finally, the ratio of FIR and radio luminosities
for the selected Seyferts and LINERs agrees with what is observed in
other analogous sources. 
Namely, the median value in our sample is
$\log(L_{\rm 60\,\mu m}/L_{\rm 5\,GHz}) \sim 5.37$ for Seyferts and
LINERs. 
These can be compared with the medians claimed
by \citet{ho01} for Seyferts, $\sim$5.31, and for regular spiral
galaxies, $\sim$5.64, implying that the Seyferts and LINERs included in
our sample are not more than twice brighter in radio than expected if
all the radio emission is due to starforming activity, in agreement
with the established properties of other Seyferts. 
We note
that the analogous ratios for the two radio galaxies included in the
sample and detected at FIR are significantly lower, $\log(L_{\rm
60\,\mu m}/L_{\rm 5\,GHz}) \sim 2.8$ and 3.5, as expected.

\begin{figure}
%\centerline{
%\includegraphics[width=3.5in,angle=270]{f3.ps}
\includegraphics[height=3.6in,angle=270]{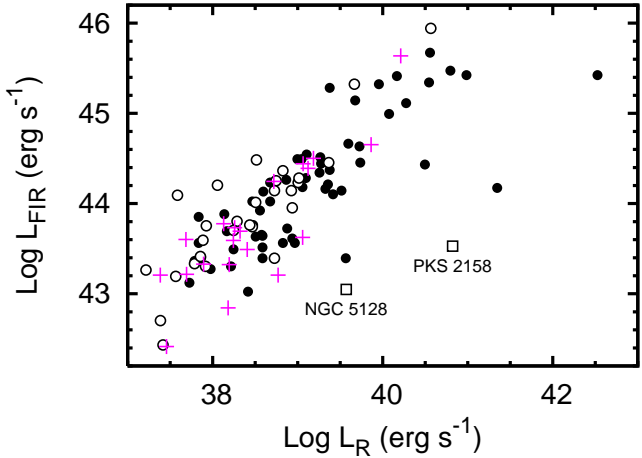}
%}
\caption{60~$\mu$m luminosity versus 5~GHz total luminosity for
Seyfert galaxies from the sample constructed by 
\citet[][detections: {\it black circles}, 
upper limits: {\it open circles}]{roy98}, 
Seyfert/LINER galaxies from our sample with the 
60~$\mu$m and 5~GHz fluxes provided ({\it magenta crosses}), 
radio galaxies from our sample with the $60$~$\mu$m fluxes 
provided ({\it open squares}). 
In the case of the sample of \citet{roy98}, sources classified as
quasars were omitted, and radio luminosities provided at 2.4~GHz were
converted to 5~GHz luminosities assuming radio spectral index $\alpha_R = 0.7$. 
} 
\label{fig:L2}
\end{figure}

\placefigure{figure~\ref{fig:L2}}

Figure~\ref{fig:L2} shows the
60~$\mu$m luminosity versus 5~GHz total luminosity for
Seyfert galaxies from the sample constructed by 
\citet{roy98}, Seyfert/LINER galaxies from our sample with the 
60~$\mu$m and 5~GHz fluxes provided, and
radio galaxies from our sample with the 60~$\mu$m fluxes 
provided. 
In the case of the sample of \citet{roy98}, sources classified as
quasars were omitted, and radio luminosities provided at 2.4~GHz were
converted to 5~GHz luminosities assuming a radio spectral 
index $\alpha_R = 0.7$. 
All our selected Seyferts/LINERs match the FIR--radio correlation 
established for other nearby Seyferts. 
Our selected radio galaxies are over-luminous
in radio for a given FIR luminosity with respect to Seyferts, as expected.

\subsection{Selected radio galaxies} \label{selected}
%##############################################################################

As
already discussed 
in \S\ref{agn_ph},
radio activity in low-power
active galaxies of the Seyfert or LINER type differ substantially from
that observed in well established jet sources like radio galaxies
and radio-loud quasars. 
Indeed, there are some important reasons for 
such a difference. 
In particular, Seyferts and LINERs are usually hosted by 
spiral (disk) galaxies, while radio galaxies and radio-loud quasars are 
typically hosted 
by giant ellipticals. 
It was noted recently that merger episodes triggering
jet activity (by shaping the accretion processes and even determining 
spins of supermassive black 
holes), 
proceed differently depending on the host properties \citep{hop06,sik07,vol07}.

\begin{figure}[t]
\centerline{
\includegraphics[height=3.2in]{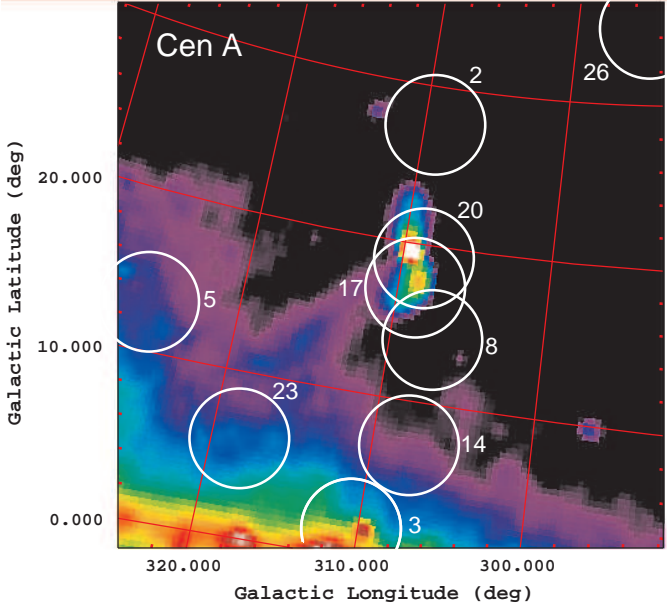}
}
\caption{
Radio map \citep[at 408 MHz from ][]{Haslam1982} of 
the $35^\circ \times 35^\circ$
field centered on the nearby radio galaxy Cen~A.
The total extent
of the north-south radio lobes is $\sim$$9^\circ$ and is centered on the
AGN (the bright white region near the center of the field). 
The $r=3.2^\circ$ circles 
mark the positions of the UHECR events detected in the field by
\PAO{} \citep{Abraham2008a}. 
The numbers correspond to the event number 
as provided in \citet{Abraham2008a}, and also in our Table~\ref{t1}.
Note event \#3 corresponds most closely to Cen~B, a bright spot 
near the center of the circle, shown with higher resolution in Figure~\ref{fig:CenB}.
}
\label{fig:CenA}
\end{figure}

\begin{figure}[t]
\centerline{
\includegraphics[height=3.2in]{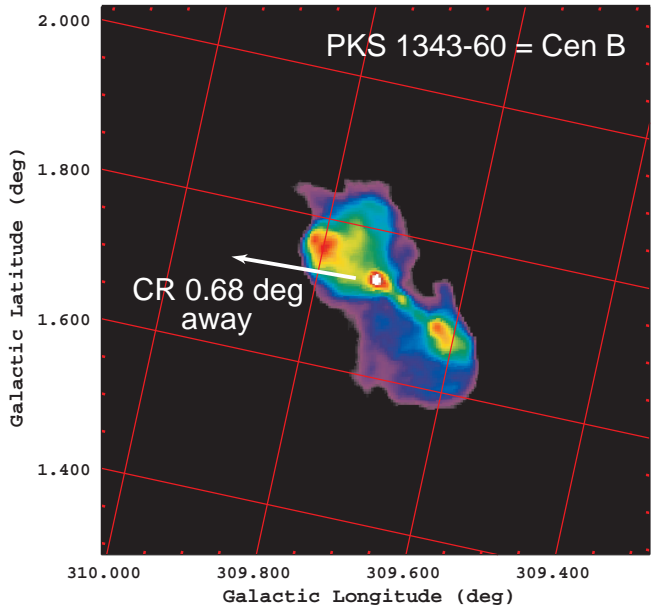}
}
\caption{
Radio image of the $0.75^\circ \times 0.75^\circ$ field centered on the 
radio galaxy 
Cen~B. 
The $43\arcsec$ resolution image was obtained with the MOST at 843 MHz
by \citet{1991PASAu...9..255M}. 
The location of the closest CR detected by \PAO{}
(\#3; c.f.\ Figure~\ref{fig:CenA}) is
indicated by the arrow pointing away from the radio nucleus.
}
\label{fig:CenB}
\end{figure}

\begin{figure}[t]
\centerline{
\includegraphics[height=3.2in]{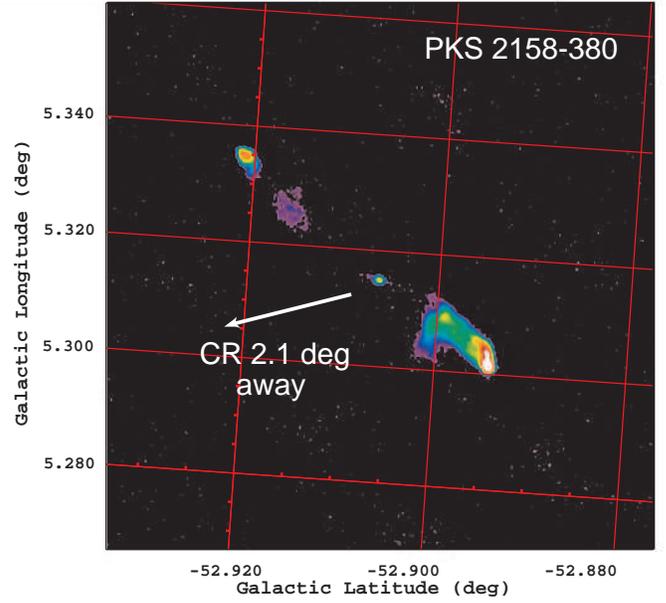}
}
\caption{
Radio image of the $200\arcsec \times 200\arcsec$ field around the radio
galaxy PKS~2158--380. This VLA 4.9 GHz image at 2$\arcsec$ resolution was made
using a multi-configuration dataset consisting of a 1 hr observation in
Dec 1983 (program AT45) and 10 min observation from Jun 1997 (AK444).
The latter dataset was obtained through the NRAO VLA Archive
Survey (NVAS). The location of the closest
cosmic ray detected by \PAO{} (\#9) is indicated by the arrow
pointing away from the radio nucleus.
(This NVAS image was produced as part of the NRAO VLA Archive Survey, (c) AUI/NRAO.)\\
}
\label{fig:PKS2158}
\end{figure}

\begin{figure}[t]
\centerline{
\includegraphics[height=3.2in]{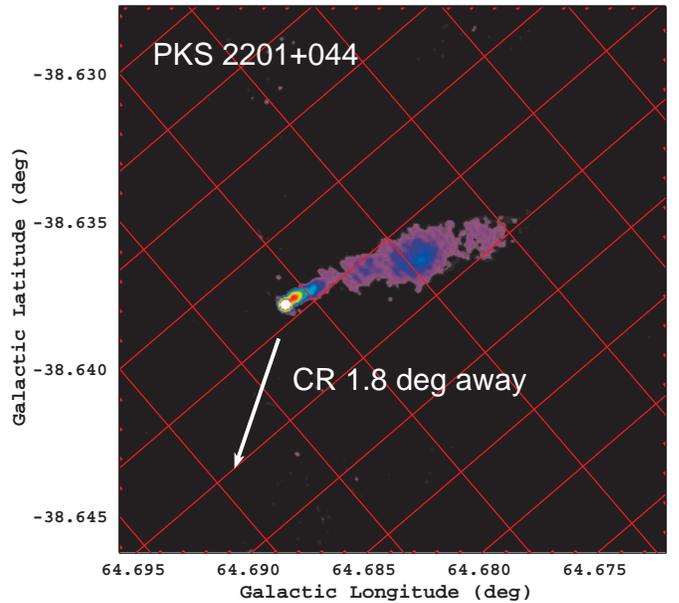}
}
\caption{
Radio image of the $100\arcsec \times 100\arcsec$ field around the 
BL Lac object PKS~2201+044 showing its one-sided radio jet. 
The image is a lower resolution ($1\arcsec$ beam) version of the VLA 8.5 GHz data 
published in \citet{2007ApJ...670...74S}.
The location of the closest CR detected by \PAO{} (\#19) is 
indicated by the arrow pointing away from the radio
nucleus.
}
\label{fig:PKS2201}
\end{figure}

Of the 54 selected AGN (Table~\ref{t2}), 3 are classified as radio galaxies and one 
is a BL Lac object, i.e., under the unification scheme, it is a FR~I 
observed with a small jet viewing angle \citep{Urry1995}. 
This group includes 
Cen~A (Figure~\ref{fig:CenA}), which is characterized by a 
well-known FR~I radio morphology with a one-sided ($\sim$4 kpc long) jet and 
giant ($\sim$0.5 Mpc) radio lobes \citep[e.g.,][]{isr98}.
This is the only
source in the sample detected in \gray{s} \citep{ste98,sre99}.
Radio maps of Cen~B
reveal a one-sided well-defined FR~I jet extending 
from pc- to kpc-scales, and an edge-brightened structure on the opposite 
lobe more characteristic of a powerful FR~II \citep[Figure~\ref{fig:CenB},][]{jon01}.
From our new VLA map of PKS~2158--380 (Figure~\ref{fig:PKS2158}), its radio morphology is 
characteristic of a FR~II with bright compact features at the outer edges 
of its lobes. It is, however, relatively underluminous for a FR~II 
($L_{\rm 1.4\,GHz}\sim 4 \times 10^{24}$ W Hz$^{-1}$) and can be considered an 
intermediate object.
PKS~2201+044 is classified as a BL Lac object with an asymmetric core-jet 
radio structure (Figure~\ref{fig:PKS2201}), and an extended ($\sim$100 kpc) radio halo 
\citep{aug98}.
These four objects are quite representative of    
the local population of radio galaxies. 
Note that Cen~A is exceptionally 
bright and extended in the sky only due to its proximity.

In fact, due to its large extent on the sky, Cen~A may be considered as
being likely associated with more than two Auger events (see Figure~\ref{fig:CenA}).
Note that Table~\ref{t1} indicates a possible association of Cen~A with 
only two CR events, because 
the $3.2^{\circ}$ separation of the UHECR event from the Cen~A \emph{nucleus} 
is assumed
when compiling the list of the selected AGN).
Considering the 9 events plotted in Figure~\ref{fig:CenA} around the giant radio 
structure of Cen~A,
their detection rate appears 
steady (almost every third event
detected by \PAO, except for the larger gap between events \#8 and \#14). 
If the rate is indeed steady, as would be expected if giant Mpc-scale radio
lobes of Cen~A are the acceleration sites of these events, this will 
be easily 
tested with 
additional 2-4 years of \PAO{} observations.

\placefigure{figure~\ref{fig:CenA}}
\placefigure{figure~\ref{fig:CenB}}
\placefigure{figure~\ref{fig:PKS2201}}
\placefigure{figure~\ref{fig:PKS2158}}

Considering the four events closest to Cen~A, 
the positions of these four events are roughly aligned with the axis of the
radio lobes, which is also aligned with the super-galactic plane 
\citep[cf.\ Figure 2 of][]{Abraham2008a}. 
The latter two events are closest to the
center of Cen~A and the former two are coincident with other AGN in the
field. 
It is of interest to note that one of the field AGN is NGC 4945
(event \#8), which is located in the Centaurus group (i.e., it is at the same
approximate distance as Cen~A). 
NGC 4945 hosts a less powerful radio source
than Cen~A and is dominated by extended emission from the galaxy and a
compact non-thermal core \citep{Elmouttie1997}.
Also, of the 6 UHECR events without an associated nearby
($z<0.018$) AGN within a $\theta\le3.2^\circ$
circular area, 2 are in the plotted
field (\#14, \#26). 
However, \#14 is the one closest to the
Galactic Plane, where it is more difficult to identify AGN,
and \#26 is the one furthest from Cen~A. 
As discussed in the next Section
a larger deflection angle for the events close to the Galactic plane 
is possible, which could mean that Cen~B could be associated with more 
than 1 event.

%\newpage
\section{Propagation and Composition of UHECRs} \label{propagation}
%##############################################################################

The propagation of the UHECRs from the sources to the observer is not 
rectilinear due to deflection by intervening magnetic fields.
The magnetic field structure (both extragalactic and Galactic), along with 
the UHECR source distribution,
the nature of sources 
(transient vs.\ steady), the energy spectrum at the injection, and the CR
composition, are all
``known unknown''
factors that affect the distribution of the observed 
arrival directions.

Though quite an extensive literature on simulation of UHECR propagation
in extragalactic magnetic fields exists,
very little is known about the strength and configuration of such fields. 
So far, direct evidence for the presence of
extragalactic magnetic fields has been found only in galaxy clusters 
\citep[for a review see][]{CT2002}.
Faraday rotation measurements provide evidence for intracluster core fields
in the range of $1-10$ $\mu$G. 
Outside clusters only upper limits at the
level $1-10$ nG are available. 
Extragalactic magnetic fields are 
\emph{ad hoc} assumed to have a domain structure with a Kolmogorov 
power spectrum and a uniform correlation length. 
Toy models assume a field strength $\sim$1 nG in voids with a somewhat larger 
field $\sim$10 nG at the supergalactic plane 
\citep[e.g.,][]{Stanev2003}. 

Recent simulations of magnetic fields in the intergalactic medium are 
more sophisticated. 
They take into account the growth of the magnetic
fields from seed fields such that the resulting field strength traces
the baryon density as the large-scale structure evolves.
A more realistic extragalactic magnetic field of this nature 
may result in significantly larger deflections
than is expected from a purely random field.
A simulation by \citet{Sigl2004} used the Biermann battery mechanism to generate
seed fields which were evolved, and then 
rescaled
so that the magnetic field in the core of a simulated 
Coma-like galaxy cluster is comparable to the $\mu$G fields 
as indicated by Faraday rotation measures.
Simulations of large-scale
structure formation and the build-up of magnetic fields in the intergalactic
medium have also been performed by \citet{Dolag2005}. 
The basic assumption
is that cosmological magnetic fields grow in a magnetohydrodynamic 
amplification
process driven by the formation of structure from a magnetic seed field
present at high redshift. 
The initial density fluctuations were
constructed from the IRAS 1.2-Jy galaxy survey by first smoothing the
observed galaxy density field on a scale of 7 Mpc, evolving it linearly
back in time and then using it as a Gaussian constraint for an otherwise
random realization of a $\Lambda$CDM cosmology \citep{Mathis2002}.
As a result, the positions and masses of prominent galaxy clusters coincide 
closely
with their real counterparts in the local universe.
\citet{Takami2006} have used a magnetic field strength scaled with 
the matter density $|B|\propto\rho^{2/3}$,  where the distribution of galaxies
is constructed using the IRAS PSCz catalog.
The correlation length is taken to be 1 Mpc
and the magnetic field is assumed to be represented as a Gaussian random
field 
with a Kolmogorov power spectrum in each cube. 
The field is further renormalized to obtain $\sim$0.4 $\mu$G in a cube
that contains the center of the Virgo Cluster.

The average deflection angle
in a random field is 
$\thetaav\approx 2.5^\circ Z E_{20}^{-1} B_{-9}({D_{100}l_1})^{1/2}$,
where $Z$ is the particle charge, $D_{100}$ is the distance in units 100 Mpc,
$B_{-9}$ is the r.m.s.\ field strength in nG, $E_{20}$ is the 
particle energy in units $10^{20}$ eV, and $l_1$ is the correlation length
in Mpc \citep[e.g.,][]{Waxman1996}.
%,Sigl2004}. 
The \emph{average} time delay corresponding to the \emph{average} deflection angle 
$\thetaav$ is $\langle\tau\rangle\sim\langle \theta_d^2\rangle D/4c$
(\citealt{Alcock1978}, their eq.~[29]), 
where $\langle \theta_d^2\rangle=4\thetaav^2/\pi$
can be derived from $\theta_d^2$ probability distribution 
(their eq.~[23]),
and $c$ is the speed of light. This yields
%The time delay corresponding to $\theta_d$ can be estimated as 
$\langle\tau\rangle\sim \thetaav^2 D/\pi c
\sim 2\times10^5 Z^2 E_{20}^{-2} B_{-9}^2 D_{100}^2 l_1$ yr. 
For a $10^{20}$ eV proton injected at $D\sim75$ Mpc for characteristic 
values of $B_{-9} \sim 1$ and $l_1 \sim 1$, $\thetaav\sim2^\circ$ and 
$\langle\tau\rangle\sim 1.6\times10^5$ yr which
is comparable to the crossing time of the Galaxy and the characteristic 
timescale of jet lifetimes in Seyfert galaxies as discussed earlier, but 
is negligible
compared to the galactic evolution timescale.
Therefore, observations at different wavelengths show us \emph{nearly}
a snapshot of the sources at the time when the highest energy CRs were 
emitted.
If a detected CR particle has been accelerated by a pc-scale jet, the 
jet will expand during the time delay to become larger, 
$\sim$$10^5$
lt-yr, and this may be seen as a more extended structure
in the radio. 
The association of UHECR accelerators must correspondingly take into account
such time delays and source evolution since observed 
photon signals come from later times than the epoch of UHECR escape from
the source.

The Galactic magnetic field is known much better than the extragalactic one.
It can be determined from pulsar rotation and dispersion
measures combined with a model for the distribution of free electrons
\citep[e.g.,][]{Cordes2003a,Cordes2003b}.
A large-scale field of a few $\mu$G aligned with the spiral arms exists, 
but there is no general agreement on the details \citep{Beck2001}. 
Recent studies give a bisymmetric model for the large-scale Galactic 
magnetic field with reversals on arm-interarm boundaries 
\citep{Han2006,Brown2007,Han2008}. 
Independent estimates of the strength and distribution of the
field can be made by simultaneous analysis of radio synchrotron, CR, and \gray{}
data, and these confirm a value of a few $\mu$G, increasing towards the 
inner Galaxy \citep[][and Strong et al. {\em in prep.}]{Strong2000}. 
The magnetic field in the halo is less known. 
Observations of the rotation measure of extragalactic radio sources 
reveal azimuth magnetic fields
in the halo with reversed directions below and above the plane consistent 
with A0 symmetry type \citep{WK1993,Han2008}.

Because of their large Larmor radii $>$1 kpc
($r_L\approx 10^5 E_{20}/B_{-9}$ kpc), UHECRs propagating in the Galactic 
magnetic field are sensitive to 
the global topology of the field. 
The influence of the geometry of the Galactic
magnetic field has been studied in various source
distribution scenarios \citep{Stanev1997,Alvarez2002,Takami2006}.
For a $\sim$1 $\mu$G magnetic field, the distance $D\sim100$ kpc, 
and a correlation length $l\sim1$ kpc,
the average deflection angle is $\thetaav\sim 2.5^\circ E_{20}^{-1}$, but
the actual value depends on the arrival direction of a CR particle.
Cen~A is only $\sim$$50^\circ$ away in 
longitude from the Galactic Center, and only $\sim$$20^\circ$ from the 
Galactic plane, while Cen~B is very close $\sim$$1^\circ$ to the Galactic 
equator.
Cosmic rays coming from either of these objects could be influenced by the
stronger magnetic field near the Galactic plane
(a few $\mu$G vs.\ $\sim$1 $\mu$G in the Galactic halo)
over tens of kpcs of their trajectory.
This would provide a greater deflection than the relatively longer path length
through the weak extragalactic magnetic field.
Therefore, an association of Cen~A and Cen~B with more events in this region
is plausible.

The UHECR source distribution is usually assumed homogeneous or to follow
the baryon density distribution. 
The former case is relevant for energies
below the photopion production threshold for proton injection 
where the energy losses are small and particles may come
from cosmological distances. 
Since only a small fraction of the sky is covered with an extragalactic 
magnetic field capable of deflecting UHECR particles by a significant angle 
\citep{Dolag2005},
the resulting distribution of arrival directions is close to 
isotropic \citep{Takami2006}. 
If the observed 
energy of CRs is near the GZK cutoff, the sources are likely local.
In this case the distribution of sources traces the baryon density distribution
in the local Universe
and the effective field acting on UHECRs should be considerably 
stronger, leading
to larger deflection angles. The distribution of the deflection angles 
depends on the details of the simulations, e.g., \citet{Sigl2004} 
predicts large deflection angles, $\sim$20$^\circ$ at $10^{20}$ eV, 
while \citet{Dolag2005} gives much smaller angles, but allows for
angles $>$$3^\circ$ in a small fraction of the sky $<$0.01.
Therefore,
the arrival directions of UHECRs \citep{Abraham2007b}
should correlate with the distribution of large deflections on a deflection 
map; such a correlation can be seen even from a by-eye comparison with 
the deflection map given by \citet[][their Figure~5]{Takami2006}.
The sources of UHECRs
should also be capable of producing lower energy CRs and 
\grays{} (and neutrinos) and, therefore, may be observed with the next 
generation \gray{} telescopes.

Most of our discussion has described the situation if the UHECRs particles 
are protons.
This is complicated further if the injected particles are CR nuclei 
since the deflection angles can be larger for a given magnetic field and the 
nuclei undergo photodisintegration processes on the CMB and 
extragalactic infrared background fragmenting into lighter nuclei 
\citep[e.g.,][]{Stecker1999}.
The UHECR chemical composition is unknown and subject to considerable debate.
Results from the surface arrays AGASA \citep{Shinozaki2006b} 
and Yakutsk \citep{Knurenko2008} and fluorescence detectors 
\citep[e.g.,][and references therein]{Sokolsky2007} indicate a trend toward 
proton dominated composition at the highest energies.
However, the \AC{}
has presented a fit 
to the elongation rate\footnote{The elongation rate is the slope 
$dX_{\max}/d\log E$, where $X_{\max}$ is the depth of shower maximum.} 
\citep{Unger2007} showing a heavier or mixed composition at the highest 
energies. 
These interpretations are complicated by the necessary reliance on 
hadronic interaction models which have to extrapolate cross section information 
beyond current accelerator energies, and indeed even details 
of the UHECR sources themselves can introduce degeneracy in the interpretation
of the data with different chemical compositions \citep[e.g.,][]{Arisaka2007}.
It has been argued recently 
\citep[e.g.,][]{Hooper2008,Fargion2008,Dermer2008} that the
current anisotropy results can be explained if the composition has a 
significant component of light nuclei, $4 \leq A \leq 14$, but this remains to 
be tested by further data. 

\section{Conclusion} \label{conclusion}
%##############################################################################

A transition from an isotropic distribution of arrival directions 
of CRs above $\sim$1 EeV \citep{Watson2008} to an anisotropic distribution of 
the highest-energy CRs above 57 EeV \citep{Abraham2007b} observed by \PAO{}
implies a change in the propagation mode of UHECRs 
in intergalactic and/or Galactic space. 
The association of the observed events with the supergalactic plane \citep{Abraham2007b,Stanev2008}
points to the sources tracing the supergalactic plane and matter distribution
which correlates with AGN. 
However, as we have shown, almost all nearby 
($d_{\rm L} \leq 150$\,Mpc) active galaxies found within the search radii of 
$3.2^{\circ}$ around the UHECR events detected by \PAO{}
are typical for the local low-luminosity AGN of the Seyfert/LINER type. 
They are characterized by low-power and short jet activity, which is 
substantially different from that observed in radio galaxies and quasars
(as typically considered in the scenarios for acceleration of UHECRs). 
Moreover, such selected low-luminosity AGN are expected to be
quite common in the local Universe, with the estimated surface 
density
$1.4\times10^3$ per steradian, when limited to a redshift $z \leq 0.037$.
If the acceleration of UHECRs is indeed associated with jet activity, which
is most likely,
we conclude that the correlation with particular AGN is a coincidence.
To distinguish between the persistent and episodic \citep[e.g.,][]{Farrar2008} models of UHECR sources, future, 
more extensive analyzes have
to take into account details of the AGN 
radio morphology and spectral properties, and may yield
a correlation with a larger deflection angle and/or more distant sources.

We emphasize that there is no complete all-sky catalog of nearby AGN. 
In addition, 
many ``regular'' galaxies when studied at sufficient spatial resolution at
different wavelengths show some (typically weak) level of the AGN-like activity.
Hence, the confusion in classification of such sources in the literature,
different databases, and catalogs. 
Thus, investigating the correlation of UHECRs
with AGN based on some given particular AGN catalog may be tricky and 
even meaningless. 
In particular, using catalogs of X-ray selected local active galaxies \citep[as in, e.g.,][]{geo08}
may give misleading results since the X-ray emission of Seyferts and 
LINERs is produced 
by the accretion disks and the disk coronae, and therefore represents 
the accretion power of the active nucleus rather than the power of its jet.
Although the most recent studies have found a correlation between the disk 
luminosity and the radio power of the unresolved nucleus \citep{ho02,hp01,pan06,pan07},
the former
has no direct relation with the large-scale radio structures which
are supposed to be capable of accelerating CRs up to the highest energies.
Besides, the \emph{present time} X-ray luminosity of the disks may have
nothing to do with the \emph{observed} UHECR events because of the 
considerable time delay between the arrivals of particles and photons (see \S\ref{propagation});
on the other hand, \emph{past} UHECR acceleration activity that produced the 
observed UHECR events, 
if not episodic,
has to manifest itself
by extended jets that we should be able to see \emph{now}.
As argued in this paper, the
spectral and 
morphological properties of the \emph{jetted} AGN which are selected as likely 
counterparts of the detected UHECR events should be considered in detail and 
compared with the properties of the parent population.

Other possibilities include a few close sources with 
extended jet/lobe
structures, such as Cen~A and Cen~B, and relatively large deflections 
due to either stronger magnetic fields or due to the presence of 
heavy nuclei in the flux, or more distant sources.

Observations with \gray{} telescopes, such as 
\emph{Fermi}/LAT,
HESS, MAGIC, and VERITAS
may point to the \emph{class of sources} able to accelerate 
particles to TeV energies,
and are therefore potentially capable of accelerating particles up to EeV 
energies.
Such sources could also produce TeV and UHE neutrinos.
Taking into account the delay between the arrival times of 
\grays{} (neutrinos) and UHECRs,
such observations have to be interpreted with care: UHECRs may come 
from sources which are not generating TeV \grays{} anymore, or UHECRs that
are accelerated in present day \gray{} emitters have not had time to propagate
to us, yet.
Meanwhile, Cen~A and Cen~B are two
powerful nearby radio galaxies 
and, if they are indeed UHECR sources, 
\gray{} observations can provide a ``current'' picture at the time when the 
CRs were emitted since the overall time delay from propagation is very short.
Moreover, Cen~A is large enough to be resolved by \gray{} instruments \citep[e.g., 
\emph{Fermi}/LAT,
][]{McEnery2004,glast}.
Therefore,
observations with \gray{} telescopes may provide additional clues to the origin of UHECRs.

%\newpage
\acknowledgments
I.\ V.\ M.\ acknowledges support from NASA
Astronomy and Physics Research and Analysis Program (APRA) grant.
\L{}.\ S. acknowledge support by the MEiN grant 1-P03D-003-29.
T.\ A.\ P.\ acknowledges partial support from the US Department of Energy.
C.C.C. was supported by an appointment to the NASA Postdoctoral
Program at Goddard Space Flight Center, administered by Oak Ridge
Associated Universities through a contract with NASA. This research has
made use of the NASA/IPAC Extragalactic Database (NED) which is operated by
the Jet Propulsion Laboratory, California Institute of Technology, under
contract with NASA.

\end{document}